\documentclass{article}
\usepackage{float}
\usepackage{amsmath}
\usepackage{amssymb}
\usepackage{graphicx}
\usepackage{tabularx}
\usepackage{booktabs}
\usepackage{multirow}
\usepackage[accepted]{icml2023}

\begin{document}

\icmltitlerunning{Generating Protein Structures by Equivariantly Diffusing Oriented Residue Clouds}

\twocolumn[
\icmltitle{Generating Novel, Designable, and Diverse Protein Structures\\ by Equivariantly Diffusing Oriented Residue Clouds}

\begin{icmlauthorlist}
\icmlauthor{Yeqing Lin}{coms,sbio}
\icmlauthor{Mohammed AlQuraishi}{coms,sbio}
\end{icmlauthorlist}

\icmlcorrespondingauthor{Mohammed AlQuraishi}{m.alquraishi@columbia.edu}

\icmlaffiliation{sbio}{Department of Systems Biology, Columbia University, New York, NY, USA}
\icmlaffiliation{coms}{Department of Computer Science, Columbia University, New York, NY, USA}

\icmlkeywords{De Novo Protein Design, Denoising Diffusion Probabilistic Models}

\vskip 0.3in
]

\printAffiliationsAndNotice{}

\begin{abstract}
Proteins power a vast array of functional processes in living cells. The capability to create new proteins with designed structures and functions would thus enable the engineering of cellular behavior and development of protein-based therapeutics and materials. Structure-based protein design aims to find structures that are designable (can be realized by a protein sequence), novel (have dissimilar geometry from natural proteins), and diverse (span a wide range of geometries). While advances in protein structure prediction have made it possible to predict structures of novel protein sequences, the combinatorially large space of sequences and structures limits the practicality of search-based methods. Generative models provide a compelling alternative, by implicitly learning the low-dimensional structure of complex data distributions. Here, we leverage recent advances in denoising diffusion probabilistic models and equivariant neural networks to develop Genie, a generative model of protein structures that performs discrete-time diffusion using a cloud of oriented reference frames in 3D space. Through \textit{in silico} evaluations, we demonstrate that Genie generates protein backbones that are more designable, novel, and diverse than existing models. This indicates that Genie is capturing key aspects of the distribution of protein structure space and facilitates protein design with high success rates. Code for generating new proteins and training new versions of Genie is available at \url{https://github.com/aqlaboratory/genie}.
\end{abstract}

\section{Introduction}
\label{introduction}

Proteins play a key role in cellular processes, ranging from chemical catalysis to molecular transport. Over the course of evolution, nature has explored a plethora of protein structures and accordant functions. Yet, relative to the potential size of foldable protein space, evolution has only explored a small subregion \cite{huang2016coming}. This suggests the possibility of designing new proteins unlike any seen in nature, if suitable methods can model uncharted parts of fold space. Protein design methods have historically focused on optimizing functional properties of natural proteins through directed evolution \cite{dougherty2009directed} or through rational design of novel protein sequences that hew closely to known structural motifs \cite{kuhlman2003design}. This limited exploration of fold space to regions adjacent to natural proteins. With recent advances in protein structure prediction methods, new approaches have been proposed that leverage representations learned by these methods to more broadly explore structure space. For example, Anishchenko et al. \yrcite{anishchenko2021novo} performed Monte Carlo sampling in sequence space using trRosetta \cite{yang2020improved} as a guide and were able to discover novel structures. One disadvantage of this approach however is the reliance on sampling, which can be computationally expensive and difficult to steer toward desirable design goals. Generative models that capture complex data distributions provide a new direction for \textit{de novo} protein design. In lieu of sampling from protein sequence space, new designs could be discovered by implicitly learning the space of structures.

\textbf{Generative modeling trilemma} \quad Generative models generally contend with a trilemma in optimizing between quality (physicality and designability of protein structures), mode coverage (novelty and diversity of structures) and sampling time \cite{xiao2022DDGAN}. Multiple modeling paradigms exist --- for example, Generative Adversarial Networks (GANs) \cite{NIPS2014_5ca3e9b1} and Variational AutoEncoders (VAEs) \cite{kingma2014auto} --- each making a different trade-off. Recently, denoising diffusion probabilistic models (DDPMs) \cite{ho2020denoising,nichol2021improved} have shown considerable promise in generating high quality 2D images, as exemplified by DALL-E 2 \cite{ramesh2022hierarchical}. DDPMs consist of a forward process that iteratively adds Gaussian noise to a sample and a reverse process that iteratively removes noise from a noisy sample. DDPMs optimize for sample quality and diversity, achieving state-of-the-art performance on both \cite{dhariwal2021diffusion} at the cost of increased sampling time.

\textbf{Application to protein design} \quad Multiple prior efforts have applied generative modeling to structure-based protein design. The first generation of methods parameterized protein geometry using inter-residue distances, leveraging the pre-existing machinery for 2D image generation. For instance, Anand and Huang \yrcite{NEURIPS2018_afa299a4} used GANs to generate pairwise distance matrices of $C_\alpha$ atoms in proteins, followed by convex optimization to reconstruct the corresponding 3D coordinates. Anand et al. \yrcite{anand2019fully} later introduced an additional refinement network to improve coordinate reconstruction. One limitation of this approach is the lack of a guarantee that generated pairwise distances are embeddable in 3D space, leading to potential inconsistencies between raw samples (in distance matrix space) and generated coordinates. Errors in distance matrices often lead to significant deterioration in structural quality \cite{eguchi2022ig}, and prevent the model from being optimized in an end-to-end fashion for final 3D geometry.

An alternate parameterization for protein structure is internal coordinates, where torsion angles between adjacent residues are used to encode 3D geometry. This approach sidesteps the embeddability problem of distance-based representations, but is overly reliant on reasoning over local geometry \cite{alquraishi2019end}. One example is FoldingDiff \cite{wu2022protein}, which performs diffusion using internal coordinates with a bidirectional transformer that iteratively denoises a sequence of torsion angles. FoldingDiff yields protein-like backbones but the majority of generated structures are predicted to not be designable when assessed using self-consistency metrics (described later).

A third approach parameterizes proteins using atomic coordinates in Cartesian space. Unlike distance-based and internal coordinate parameterizations, this approach is not inherently invariant to rotations and translations (SE(3)-invariance). As proteins do not have preferred orientations or locations, capturing these invariances in a model would improve its data efficiency. Recent developments in geometric neural networks, including EGNN \cite{satorras2021n} and GVP \cite{jing2020learning}, provide powerful tools for geometric reasoning in an SE(3)-equivariant manner. Employing EGNNs for this purpose, Trippe et al. \yrcite{trippe2022diffusion} developed ProtDiff, a DDPM that directly generates the $C_\alpha$ coordinates of protein structures. Although a promising approach, ProtDiff struggles to produce geometries with realizable protein sequences. One potential reason for this is the reflection-invariant property of EGNNs, which is non-physical and frequently yields left-handed alpha helices, an exceedingly rare structural element in real proteins.

In protein structure prediction, AlphaFold2 \cite{AlphaFold2021} achieved great success by combining implicit reasoning in a latent space (evoformer module) with geometric reasoning in Cartesian space (structure module). A key feature of the latter is Invariant Point Attention (IPA), a mechanism for computationally-efficient, SE(3)-equivariant reasoning that is sensitive to reflections. IPA parameterizes proteins using rigid body frames anchored at residues, which can be defined in a consistent manner irrespective of global position or orientation by taking advantage of the polymeric nature of proteins. Using a cloud of reference frames, instead of a point cloud, retains angular information between residues and thus accounts for chemical chirality. Reformulating this construction for protein design, Anand and Achim \yrcite{anand2022protein} combine IPA with a DDPM to generate protein structures and their corresponding sequences by conditioning on secondary structure topology. They perform diffusion in frame space, using random Gaussian-distributed 3D vectors and random rotation matrices as noise (the latter does not satisfy the Gaussian assumption of the original DDPM construction). Using this approach their model shows promising empirical results for secondary structure-conditioned protein design.

In this work, we similarly combine aspects of the SE(3)-equivariant reasoning machinery of IPA with DDPMs to create an (unconditional) diffusion process over protein structures. Unlike Anand and Achim \yrcite{anand2022protein}, we introduce a geometric asymmetry in how protein residues are represented---as point clouds in the forward process (the noising procedure) and as a cloud of reference frames in the reverse process (sample generation). This yields a simple and cheap process for noising structures while retaining the full expressivity of IPA during generation, without violating the Gaussian assumption of DDPMs. The resulting model, Genie, generates diverse, designable, and novel structures. When compared to other methods, Genie achieves state-of-the-art performance on key design metrics. Nearly contemporaneous with this work, three other methods reported performant DDPMs for protein design inspired by similar ideas \cite{Ingraham2022.12.01.518682,Watson2022.12.09.519842,yim2023se}, although their architectural details and training procedures are distinct from Genie's.

\section{Methods}
\label{methods}

Genie is a DDPM that generates protein backbones as a sequence of $C_\alpha$ atomic coordinates. It performs diffusion directly in Cartesian space and uses an SE(3)-equivariant denoiser that reasons over a cloud of reference frames to predict noise displacements at each diffusion step. In Section \ref{sec2.1}, we describe our implementation of DDPMs for protein generation. In Section \ref{sec2.2}, we provide details on the SE(3)-equivariant denoiser. In Sections \ref{sec2.3} and \ref{sec2.4}, we describe how we train and sample from the model.

\subsection{Denoising Diffusion Probabilistic Model} \label{sec2.1}

Let $\mathbf{x} = [\mathbf{x}^1, \mathbf{x}^2, \cdots, \mathbf{x}^N]$ denote a sequence of $C_\alpha$ coordinates of length $N$, corresponding to a protein with $N$ residues. Given a sample $\mathbf{x}_0$ from the unknown data distribution over protein structures, the forward process iteratively adds isotropic Gaussian noise to the sample following a cosine variance schedule $\pmb{\beta} = [\beta_1, \beta_2, \cdots, \beta_T]$, where the total number of diffusion steps $T$ is set to 1,000:
\begin{equation}
    q(\mathbf{x}_t | \mathbf{x}_{t-1}) = \mathcal{N}(\mathbf{x}_{t} \mid \sqrt{1 - \beta_t} \mathbf{x}_{t-1}, \beta_t \mathbf{I})
\end{equation}
By reparameterization, we have
\begin{equation} \label{eq:2}
    q(\mathbf{x}_t | \mathbf{x}_0) = \mathcal{N}(\mathbf{x}_t \mid \sqrt{\bar{\alpha}_t} \mathbf{x}_0, (1 - \bar{\alpha}_t) \mathbf{I})
\end{equation}
where
\[ \bar{\alpha}_t = \prod_{s=1}^{t} \alpha_s \quad \text{and} \quad \alpha_t = 1 - \beta_t \]

Since the isotropic Gaussian noise added at each diffusion step is small, the corresponding reverse process could be modeled with a Gaussian distribution:
\begin{equation} \label{eq:4}
    p(\mathbf{x}_{t-1} | \mathbf{x}_t) = \mathcal{N}(\mathbf{x}_{t-1} \mid \pmb{\mu}_{\theta}(\mathbf{x}_t, t), \pmb{\Sigma}_{\theta}(\mathbf{x}_t, t) \mathbf{I})
\end{equation}
where
\[ \pmb{\mu}_{\theta}(\mathbf{x}_t, t) = \frac{1}{\sqrt{\alpha_t}} \left( \mathbf{x}_t - \frac{\beta_t}{\sqrt{1 - \bar{\alpha}_t}} \pmb{\epsilon}_{\theta}(\mathbf{x}_t, t) \right) \]
\[ \pmb{\Sigma}_{\theta}(\mathbf{x}_t, t) = \beta_t \]
By starting the reverse process from white noise and then iteratively removing noise, Genie generates new proteins (Appendix \ref{appendix_a0}). This reverse process requires evaluating $\pmb{\epsilon}_{\theta}(\mathbf{x}_t, t)$, which predicts the noise added at time step $t$. We do this using a noise predictor that forms the core of Genie.

\subsection{Noise Prediction} \label{sec2.2}

\begin{figure*}[ht]
\begin{center}
\centerline{\includegraphics[width=\textwidth]{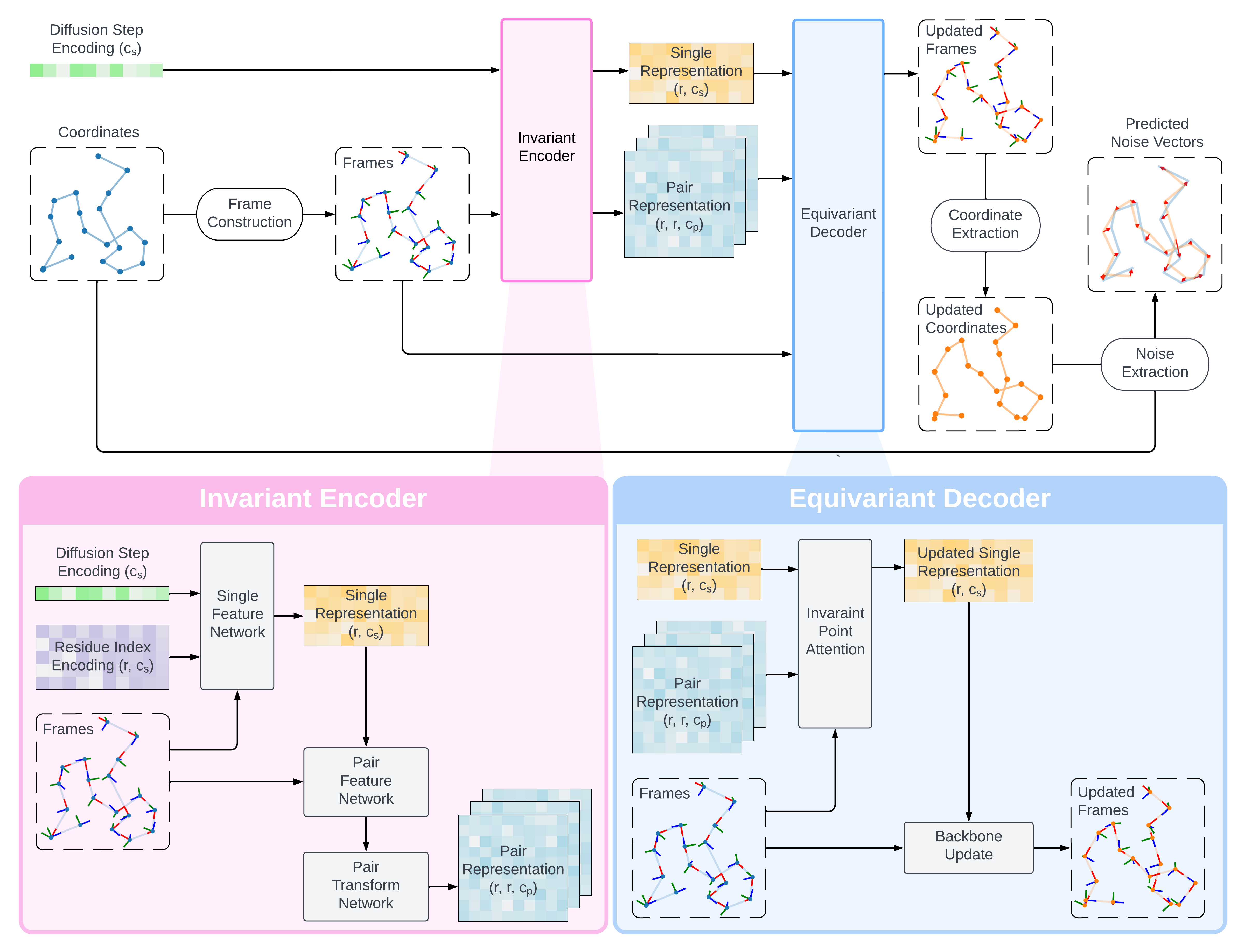}}
\end{center}
\vspace{-0.5cm}
\caption{Architecture of SE(3)-equivariant denoiser, including SE(3)-invariant encoder (bottom left) and SE(3)-equivariant decoder (bottom right). Notation: $r$: number of residues, $c_s$: dimensionality of single representation, $c_p$: dimensionality of pair representation.}
\label{architecture}
\end{figure*}

In Genie, the noise predictor first takes the $C_\alpha$ coordinates at diffusion step $t$, denoted by $\mathbf{x}_t$, and computes discrete Frenet-Serret (FS) frames based on the backbone geometry encoded by $\mathbf{x}_t$. Each FS frame represents the position and orientation of a residue relative to the global reference frame. Once constructed, these FS frames enable downstream model components, including IPA, to reason about the relative orientations of protein residues and parts. FS frames are passed together with a sinusoidal encoding of diffusion step $t$ to an SE(3)-invariant encoder and an SE(3)-equivariant decoder to compute a new set of FS frames, from which updated coordinates are extracted (see Appendix \ref{appendix_a1} for more details). Noise is then computed as a set of displacement vectors between the original and updated coordinates, which is the final prediction of $\pmb{\epsilon}_{\theta}(\mathbf{x}_t, t)$. Figure \ref{architecture} summarizes Genie's architecture.

\textbf{FS frames} \quad Following Hu et al. \yrcite{hu2011discrete} and Chowdhury et al. \yrcite{chowdhury2022single}, we construct discrete FS frames $\mathbf{F}$ as
\begin{equation*}
  \begin{split}
    \mathbf{t}^i &= \frac{\mathbf{x}^{i+1} - \mathbf{x}^i}{\lVert \mathbf{x}^{i+1} - \mathbf{x}^i \rVert}\\
    \mathbf{b}^i &= \frac{\mathbf{t}^{i-1} \times \mathbf{t}^i}{\lVert \mathbf{t}^{i-1} \times \mathbf{t}^i \rVert}\\
  \end{split}
\quad\quad\quad
  \begin{split}
   \mathbf{n}^i &= \mathbf{b}^i \times \mathbf{t}^i \\
    \mathbf{R}^i &= [\mathbf{t}^i, \mathbf{b}^i, \mathbf{n}^i]\\
    \mathbf{F}^i &= (\mathbf{R}^i, \mathbf{x}^i)
  \end{split}
\end{equation*}
% \[ \mathbf{t}^i = \frac{\mathbf{x}^{i+1} - \mathbf{x}^i}{\lVert \mathbf{x}^{i+1} - \mathbf{x}^i \rVert} \]
% \[ \mathbf{b}^i = \frac{\mathbf{t}^{i-1} \times \mathbf{t}^i}{\lVert \mathbf{t}^{i-1} \times \mathbf{t}^i \rVert} \]
% \[ \mathbf{n}^i = \mathbf{b}^i \times \mathbf{t}^i \]
% \[ \mathbf{R}^i = [\mathbf{t}^i, \mathbf{b}^i, \mathbf{n}^i] \]
% \[ \mathbf{F}^i = (\mathbf{R}^i, \mathbf{x}^i) \]
where the first element of $\mathbf{F}^i$ is the rotation matrix and the second element is the translation vector. To handle the edge cases corresponding to the N- and C-termini of proteins, we assign the frames of the second and second-to-last residues to the first and last residues, respectively.

\textbf{SE(3)-invariant encoder} \quad Given frames $\mathbf{F}_t$ and the sinusoidal encoding of the corresponding diffusion step $t$, the encoder generates and refines single residue and paired residue-residue representations, which are used later by the decoder to update the structure. As illustrated in Figure \ref{architecture} (``Invariant Encoder''), the Single Feature Network first creates per residue representations ($\mathbf{s}_t$) from sinusoidal encodings of residue indices and the diffusion step. The Pair Feature Network then computes paired residue-residue representations ($\mathbf{p}_t$) from the outer sum of the (single) residue representations, relative positional encodings of residue pairs, and a pairwise distance matrix representation of the structure (based on $C_\alpha$ coordinates). These pair representations $\mathbf{p}_t$ are iteratively refined in the Pair Transform Network using triangular multiplicative updates \cite{AlphaFold2021}. The encoder is SE(3)-invariant since both its single and pair representations are derived from SE(3)-invariant features. Appendix \ref{appendix_a2} further elaborates the encoder.

\textbf{SE(3)-equivariant decoder} \quad Given frames $\mathbf{F}_t$ and the single ($\mathbf{s}_t$) and pair representations ($\mathbf{p}_t$) from the encoder, the decoder iteratively refines the structure by operating over $\mathbf{F}_t$ in an SE(3)-equivariant manner. As illustrated in Figure \ref{architecture}  (``Equivariant Decoder''), the decoder first uses IPA to generate a new single representation $\mathbf{s}'_{t}$ based on $\mathbf{F}_t$, $\mathbf{s}_t$, and $\mathbf{p}_t$. Here, frames are initialized using $\mathbf{F}_t$ in lieu of the ``black hole'' initialization used by AlphaFold2. The Backbone Update Network then computes and applies frame updates based on the updated single representation $\mathbf{s}'_{t}$, resulting in a new set of frames $\mathbf{F}'_{t}$. The decoder is SE(3)-equivariant since frame updates are computed based on the SE(3)-invariant $\mathbf{s}'_{t}$. Thus, any global transformation of the input frames is also applied to the final output frames.

\textbf{Noise prediction} \quad Given the input coordinates $\mathbf{x}_t$ and the updated frames $\mathbf{F}'_{t}$, we extract the updated coordinates $\mathbf{x}'_t$ from the translation component of $\mathbf{F}'_{t}$ and compute the predicted noise $\pmb{\epsilon}_{\theta}(\mathbf{x}_t, t)$ as $\mathbf{x}_t - \mathbf{x}'_{t}$.

\subsection{Training}
\label{sec2.3}

Since the forward diffusion process is predefined with a fixed variance schedule, training Genie reduces to training the noise prediction model. By minimizing the error in noise prediction for each diffusion step, Genie learns to iteratively reverse the forward process and generate structures. Appendix \ref{appendix_a5} further elaborates the training process.

\textbf{Loss} \quad Following Ho et al. \yrcite{ho2020denoising}, which found that diffusion models achieve better performance when using noise $\pmb{\epsilon}_t$ as the prediction target instead of the mean in the reverse probability distribution $p(\mathbf{x}_{t-1} | \mathbf{x}_t)$, we define our loss as:
\begin{align*}
    L &= \mathbb{E}_{t, \mathbf{x}_0, \pmb{\epsilon}} \left[ \sum_{i=1}^N \lVert \pmb{\epsilon}_t - \pmb{\epsilon}_{\theta}(\mathbf{x}_t, t) \rVert^2 \right] \\
    &= \mathbb{E}_{t, \mathbf{x}_0, \pmb{\epsilon}} \left[ \sum_{i=1}^N \lVert \pmb{\epsilon}_t - \pmb{\epsilon}_{\theta}(\sqrt{\bar{\alpha}_t} \mathbf{x}_0 + \sqrt{1 - \bar{\alpha}_t} \pmb{\epsilon}_t, t) \rVert^2 \right]
\end{align*}
At each training step, we sample a protein domain $x_0$ from the training dataset, a diffusion step $t$ from a uniform distribution of integers between $1$ and $T$, and noise vectors $\pmb{\epsilon}_t$ from a unit Gaussian, and update model weights in the direction of minimizing the sum of per residue $L_2$ distances between true and predicted noise vectors.

\textbf{Datasets} \quad We employ two datasets for training and assessing models, one based on the Structural Classification of Proteins - extended (SCOPe) database \cite{fox2014scope,chandonia2022scope} and one based on the AlphaFold-Swissprot database \cite{AlphaFold2021,varadi2022alphafold}. Details of experimental setup are noted in Section \ref{sec3}. For SCOPe, we filter protein domains so that no two share \textgreater$40\%$ sequence identity, ensuring non-redundancy and diversity. We also use the SCOPe structural hierarchy to delineate domains along four major classes: all alpha, all beta, alpha and beta ($\alpha/\beta$), and alpha and beta ($\alpha + \beta$). We remove domains with multiple chains and missing backbone atoms. Our resulting training set comprises 8,766 domains, with 3,942 domains having at most 128 residues and 7,249 domains having at most 256 residues.

For AlphaFold-SwissProt, we remove structures with low confidence scores based on the predicted local distance difference test score averaged across all residues (pLDDT). pLDDT is an AlphaFold2-derived score that summarizes its own confidence in its predictions, ranging in value from 0 to 100 (higher is better). We use a cutoff of $\text{pLDDT} > 80$. The resulting training set contains 195,214 protein structures having at most 256 residues.

\subsection{Sampling}
\label{sec2.4}

To generate a new protein backbone of length $N$, we first sample a random sequence $\mathbf{x}_T = [\mathbf{x}_T^1, \mathbf{x}_T^2, \cdots, \mathbf{x}_T^N]$ of $C_\alpha$ coordinates drawn from $\mathbf{x}_T^i \sim \mathcal{N}(\mathbf{0}, \mathbf{I})$ for all $i \in [1, N]$. This sequence of coordinates $\mathbf{x}_T$ is then recursively fed through the reverse diffusion process until diffusion step $0$ is reached. Using Equation \ref{eq:4}, the update rule is:
\[ \mathbf{x}_{t-1} = \begin{cases}
    \pmb{\mu}_{\theta}(\mathbf{x}_t, t) + \sqrt{\pmb{\Sigma}_{\theta}(\mathbf{x}_t, t)} \cdot \pmb{\epsilon}, &\text{if } t > 1\\
    \pmb{\mu}_{\theta}(\mathbf{x}_t, t), &\text{otherwise}\\
\end{cases}
\]
where $\pmb{\epsilon} = [\pmb{\epsilon}^1, \pmb{\epsilon}^2, \cdots, \pmb{\epsilon}^N]$ and each $\pmb{\epsilon}^i \sim \mathcal{N}(\mathbf{0}, \mathbf{I})$.

\section{Results}
\label{sec3}

\begin{figure*}[h!]
\begin{center}
\centerline{\includegraphics[width=\textwidth]{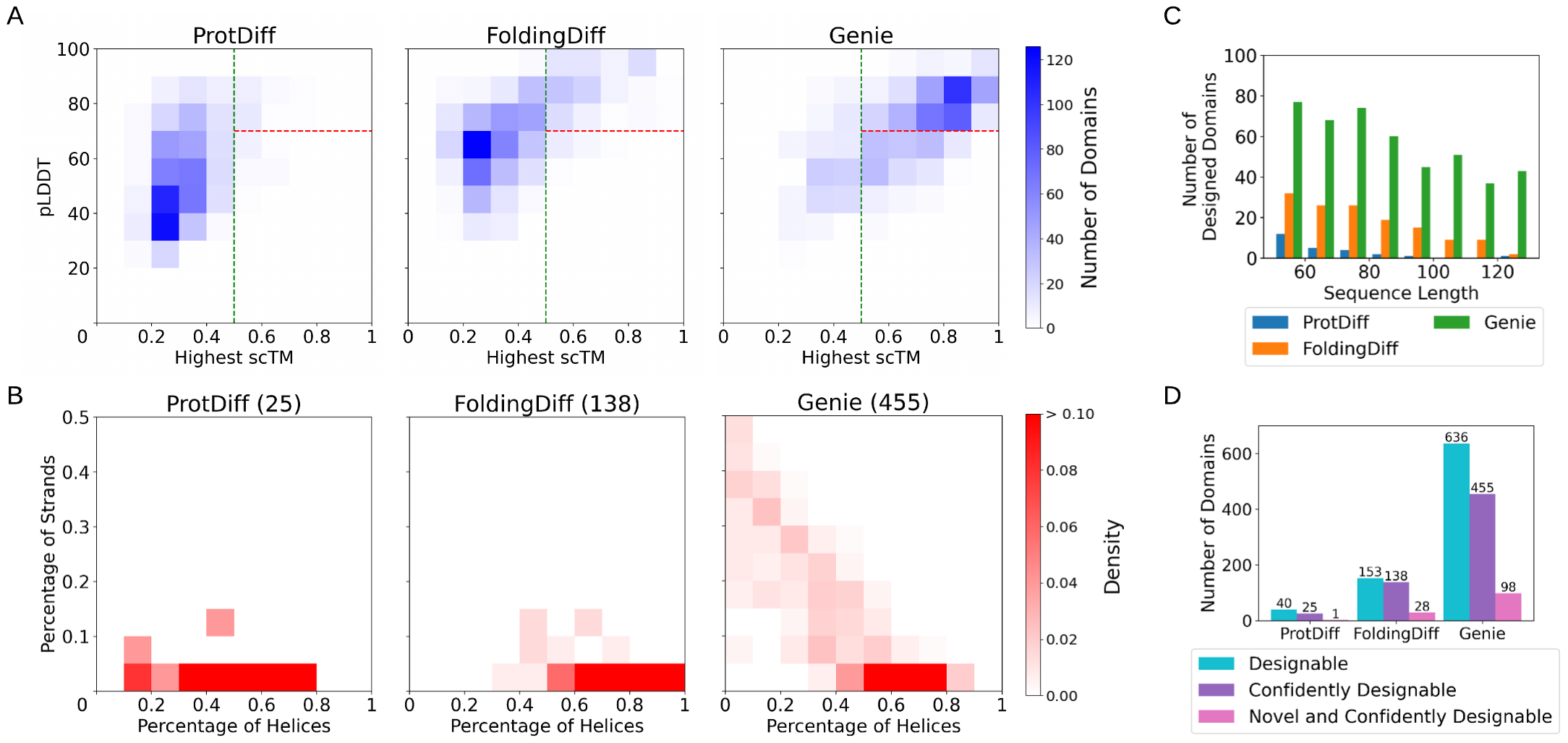}}
\vspace{-0.2cm}
\caption{Analysis of structures generated by Genie and short models. (A) Heatmap of the relative frequencies of generated domains with specific combinations of highest scTM and pLDDT values achieved by ProtDiff, FoldingDiff, and Genie. (B) Heatmap of relative frequencies of confidently designable domains with specific combinations of fractional SSE content. The number of designed domains for each model is shown in parentheses. Heatmap of relative frequencies of the SCOPe dataset is provided in Figure \ref{appendix_results}C (Appendix \ref{appendix_c1}) for reference. (C) Histogram of confidently designable domains as a function of sequence length. (D) Bar chart of number of designable domains generated by different methods out of a fixed budget of 780 attempted designs per method. }
\label{results}
\end{center}
\vskip -0.2in
\end{figure*}

To evaluate Genie, we perform two set of experiments. In Section \ref{sec3.1}, we compare Genie with ProtDiff and FoldingDiff, both of which are capable of generating proteins up to 128 residues in length (hereafter referred to as ``short" models). In Section \ref{sec3.2}, we compare Genie with FrameDiff and RFDiffusion, which are capable of generating longer proteins (hereafter referred to as ``long" models). We also visualize the design space of Genie in Section \ref{sec3.3}.

\subsection{Comparisons with short models} \label{sec3.1}

To compare with ProtDiff and FoldingDiff, we train Genie on the SCOPe dataset with maximum sequence length set to 128. To ensure fairness, we retrain ProtDiff and FoldingDiff on our filtered SCOPe dataset. For FoldingDiff, we ultimately reverted to using  the original weights as it achieves better generative performance that way (see Appendix \ref{appendix_a6} for more details). To evaluate a method, we generate 10 proteins for each sequence length between 50 and 128 residues and assess the resulting structures on designability, diversity, and novelty. We find that Genie outperforms ProtDiff and FoldingDiff on all three criteria.

\subsubsection{Designability} \label{sec3.1.1}

The first assessment criterion we consider is designability, \textit{i.e.,} whether a generated structure can be realized by a protein sequence. We follow the self-consistency Template Modeling (scTM) approach proposed by Trippe et al. \yrcite{trippe2022diffusion}. Although purely an \textit{in silico} method, it has shown promise in correctly identifying designable structures \cite{dauparas2022robust} by focusing on their overall, coarse geometry, which is suitable for the short models we consider here. Briefly, scTM takes a generated structure and feeds it into ProteinMPNN, a state-of-the-art structure-conditioned sequence generation method. Using ProteinMPNN set at a sampling temperature of 0.1, we generate eight sequences per input structure and then use OmegaFold \cite{wu2022high} to predict the structure of each putative sequence. The original scTM approach used AlphaFold2 but we substitute OmegaFold for AlphaFold2 as it outperforms the latter on single-sequence structure prediction (we also employ ESMFold \cite{lin2022language} for the same purpose and observe similar trends --- see Appendix \ref{appendix_c1}). Finally, we compute scTM by measuring the TM-score \cite{zhang2004scoring} --- a metric of structural congruence --- of the OmegaFold-predicted structure with respect to the original generated structure. scTM scores range from 0 to 1, with higher numbers corresponding to increased likelihoods that an input structure is designable. Appendix \ref{appendix_b1} illustrates this pipeline.

For each structure generated by each method, we compute the highest scTM score achieved across the eight putative sequences and the OmegaFold-derived pLDDT score for the designed structure with the highest scTM score. Figure \ref{results}A shows the distribution of highest scTM scores versus pLDDTs for all three models. Similar to previous work \cite{trippe2022diffusion, wu2022protein}, we first use $\text{scTM} > 0.5$ as a cutoff for designability since it suggests that the generated and designed structures have the same fold \cite{xu2010significant}. $81.5\%$ of protein domains generated by Genie have $\text{scTM} > 0.5$, far exceeding the percentages for ProtDiff ($5.1\%$ and $11.8\%$ for retrained and reported models, respectively) and FoldingDiff ($19.6\%$ and $22.7\%$ for resampled and reported results, respectively). Thus on average Genie yields more designable structures.

While scTM reflects a model's ability to find structures with designable sequences, it leaves open the possibility that OmegaFold-predicted structures are of insufficient quality to be used reliably in computing scTM scores. We thus place an additional constraint that predicted structures achieve $\text{pLDDT} > 70$ to enrich for confidently-predicted structures. When considering both criteria, $58.3\%$ of domains generated by Genie are designable with confidently-predicted structures (henceforth, ``confidently designable''), while only $3.2\%$ and $17.7\%$ of ProtDiff- and FoldingDiff-generated domains are, respectively. Figure \ref{results}C shows the distribution of confidently designed structures binned by sequence length. We observe that Genie outperforms ProtDiff and FoldingDiff across short and long proteins. Furthermore, Genie-generated structures universally satisfy physical chirality constraints while those generated by ProtDiff often contain left-handed helices. 

\subsubsection{Diversity} \label{sec3.1.2}

The second assessment criterion we consider is the diversity of generated structures. We first evaluate diversity by considering the relative proportion of secondary structure elements (SSEs) in generated domains. SSEs are local patterns of structure within proteins that are characterized by specific types of hydrogen bonding networks. The most common types of SSEs are $\alpha$-helices and $\beta$-strands, and we focus on these in our assessments.

To identify SSEs in generated structures, we use the Protein Secondary Element Assignment (P-SEA) algorithm \cite{labesse1997p}. P-SEA detects SSEs using a set of hand-crafted rules based on distances and angles between consecutive $C_\alpha$ atoms in protein backbones. We applied P-SEA to all confidently designable structures ($\text{scTM} > 0.5$; $\text{pLDDT} > 70$). Figure \ref{results}B shows the relative frequencies of designed domains with different fractions of SSEs. Domains generated by FoldingDiff and ProtDiff are dominated by mainly $\alpha$-helical domains, with only 2 (out of 25, $8\%$) and 10 (out of 138, $7.25\%$) of their designs containing $\beta$-strands, respectively. In contrast, Genie designs are more diverse, with 254 mainly $\alpha$-helical, 25 mainly $\beta$-strand, and 176 $\alpha,\beta$-mixed domains. 

In addition to SSE content, we assess the diversity of tertiary structures in confidently designed domains. For each domain, we compute its maximum TM score to all other confidently designed domains, which quantifies its similarity to the most structurally similar domain in the designed set. For a diverse set of domains, most domains should have small maximum TM scores to all other domains. Genie achieves, on average, a maximum TM score of $0.561 \pm 0.086$ relative to the designed set, which is lower than both ProtDiff ($0.583 \pm 0.115$) and FoldingDiff ($0.668 \pm 0.178$). This suggests that Genie-designed domains are more diverse and better able to capture the fold distribution of protein structure space.

\subsubsection{Novelty} \label{sec3.1.3}

The third assessment criterion we consider is the novelty of generated structures. As one goal of protein design is the creation of new protein folds and geometries, novelty is a key feature of any structure-based protein design tool. To quantify the novelty of generated structures we compute their maximum TM scores with respect to all structures in the training set. We use the TM-Align software package for this purpose \cite{zhang2004scoring}. To classify a confidently designable domain as novel, we require that its maximum TM score to the training set is less than 0.5 --- a widely used heuristic for determining when two protein domains are of dissimilar folds. Using this criterion, we find that 98 out of 455 ($21.5\%$) confidently designable structures generated by Genie are novel, relative to $4\%$ (1 out of 25) and $20.3\%$ (28 out of 138) for ProtDiff and FoldingDiff, respectively. Figure \ref{results}D summarizes the statistics on generated domains for all three models.

\subsection{Comparisons with long models} \label{sec3.2}

\begin{table*}[ht]
\caption{Comparison of Genie with long models on designability, diversity, and novelty of protein structure generation. Designability is computed as the number of confidently designable structures over total number of generated structures. Diversity is computed as the number of clusters over total number of generated structures, using single-linkage hierarchical clustering with a TM score cutoff of 0.6. Novelty is computed as the number of novel structures over total number of generated structures. For sampling time profiling (reported in minutes), $T$ denotes the total number of diffusion steps in the sampling process and $L$ denotes the number of residues in the structure. The performance of Genie-SCOPe on sampling time is omitted since it is in theory identical to that of Genie-SwissProt.}
\label{table_results}
\vskip 0.15in
\begin{center}
\begin{sc}
\begin{small}
\begin{tabular}{lccccccr}
\toprule
\multirow{2}{*}{Method} & \multirow{2}{*}{$N_{\text{params}}$} & \multicolumn{4}{c}{\textbf{Sampling Metrics}} & \multicolumn{2}{c}{\textbf{Sampling Time}} \\
\cmidrule(lr){3-6}
\cmidrule(lr){7-8}
 & & Designability & Diversity & $F_1$ & Novelty & $T$ & $L = 200$ \\
\midrule
FrameDiff & 17.4M & 0.483 & 0.590 & 0.531 & 0.006 & 500 & 0.388 \\
RFDiffusion & 59.8M & 0.951 & 0.667 & 0.784 & 0.170 & 50 & 0.740 \\
Genie-SCOPe & 4.1M & 0.586 & 0.738 & 0.654 & 0.039 \\
Genie-SwissProt & 4.1M & 0.790 & 0.642 & 0.708 & 0.041 & 1000 & 1.331 \\
\bottomrule
\end{tabular}
\end{small}
\end{sc}
\end{center}
\vskip -0.1in
\end{table*}

Contemporaneous with the development of Genie, new protein diffusion models have been developed that are capable of generating long proteins at high levels of quality and diversity. RFDiffusion \cite{Watson2022.12.09.519842} performs diffusion in SE(3) space by finetuning the RoseTTAFold \cite{baek2021accurate} structure prediction model to denoise at each diffusion time step. Chroma \cite{Ingraham2022.12.01.518682} utilizes a correlated Gaussian diffusion process with a covariance model to enforce protein chain and radius of gyration statistics. To reverse the added noise, Chroma utilizes random GNNs to reason over pairwise constraints, followed by equivariant structure updates via convex optimizations. FrameDiff \cite{yim2023se} uses a geometrically-principled approach for SE(3) diffusion by accounting for both translational and rotational noise in the diffusion process, and does so by training a model from scratch without relying on a pretrained structure prediction method, similar to Genie and Chroma. In this section, we restrict our comparisons to RFDiffusion and FrameDiff as the source code for Chroma is not publicly available.

To compare Genie with long models, we develop two variants of it, one trained on the SCOPe dataset and the other on the SwissProt dataset; we term these two models Genie-SCOPe and Genie-SwissProt, respectively. For both models, we use the same Genie architecture as before but set the maximum sequence length to 256. Following \citeauthor{Watson2022.12.09.519842} \yrcite{Watson2022.12.09.519842} and \citeauthor{yim2023se} \yrcite{yim2023se}, we adjust the noise scale at sampling time by multiplying the sampled noise vector with a constant factor $\eta$, analogous to low-temperature sampling. We analyze the effect of the sampling noise scale on generative performance and summarize our results in Appendix \ref{appendix_d1}. Here, we focus on the default (best-performing) sampling noise scale choices used in the original versions of the models (1 for RFDiffusion and 0.1 for FrameDiff). For all methods, we generate 5 proteins for each sequence length between 50 and 256 residues and evaluate them on designability, diversity, and novelty. We find that Genie achieves competitive performance.

\subsubsection{Designability} \label{sec3.2.1}

To assess the designability of long models, we use self-consistency Root Mean Square Deviation (scRMSD) which employs the same approach as scTM but uses the RMSD between generated and designed structures. scRMSD is more stringent than scTM due to the sensitivity of RMSD to small structural differences and is thus more suitable for assessing the more capable long models (lower scRMSD corresponds to higher quality.) Appendix \ref{appendix_b2} discusses the differences between scTM and scRMSD. We define sampling quality as the percentage of generated structures that are confidently designable ($\text{scRMSD} < 2$ and $\text{pLDDT} > 70$). Here, we compute scRMSD using the ProteinMPNN-ESMFold pipeline to maintain consistency with RFDiffusion and FrameDiff, which do the same. Table \ref{table_results} (``Designability") summarizes the percentage of generated structures that are confidently designable. We find that Genie outperforms FrameDiff in sampling quality, even when trained on the smaller SCOPe dataset, but underperforms RFDiffusion. Increasing training set size markedly improves Genie's sampling quality but RFDiffusion retains an edge. Appendix \ref{appendix_d2} provides visualizations of the distribution of scRMSDs versus pLDDTs.

\subsubsection{Diversity} \label{sec3.2.2}

\begin{figure}[h!]
\begin{center}
\centerline{\includegraphics[width=\linewidth]{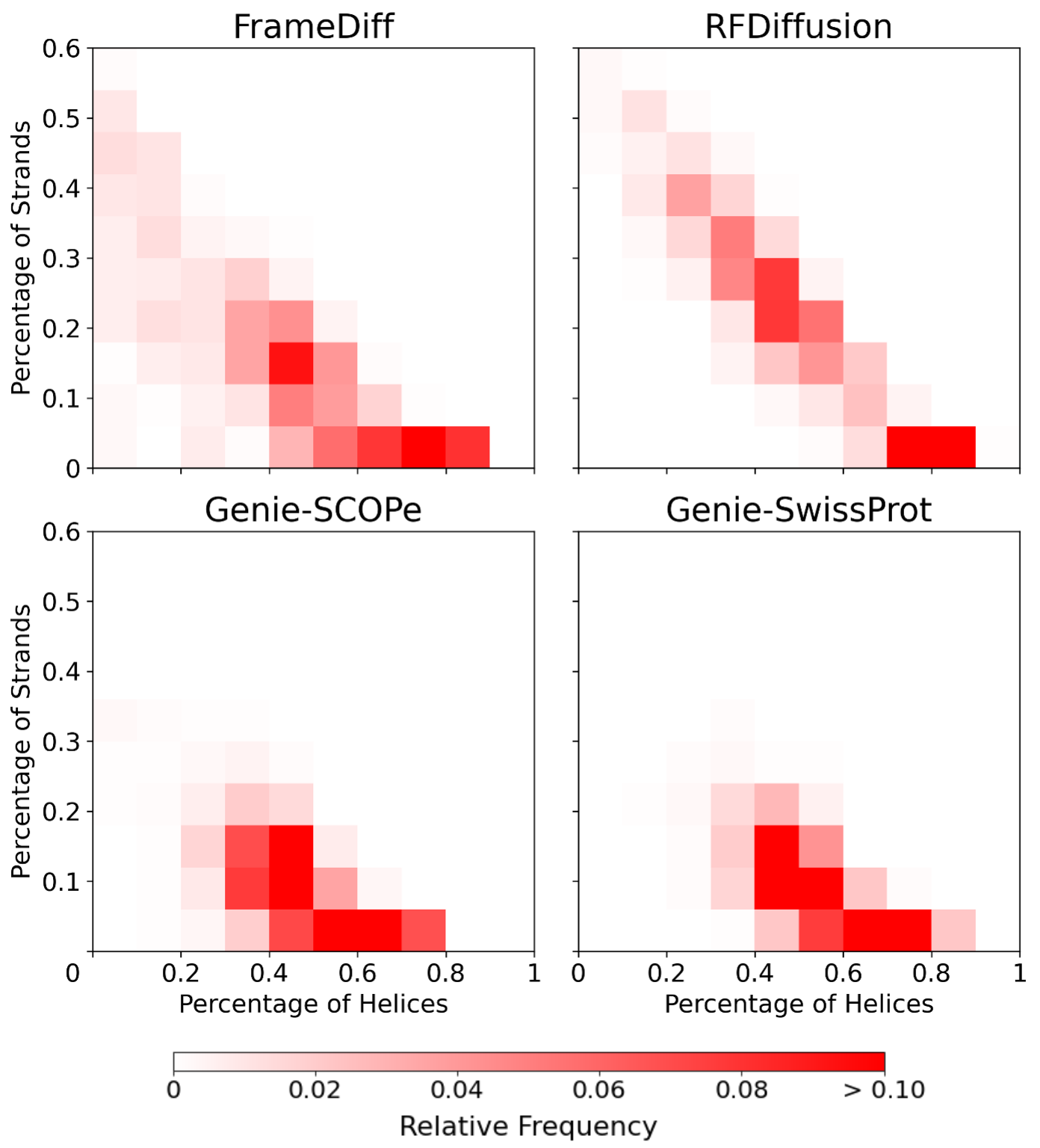}}
\vspace{-0.3cm}
\caption{Heatmap of relative frequencies of confidently designable structures with specific combinations of fractional SSE content for Genie and long models.}
\label{results_256_sse}
\end{center}
\vskip -0.2in
\end{figure}

To quantify sampling diversity, we first visualize the distribution of SSE content in Figure \ref{results_256_sse} (similar to Figure \ref{results}B in the short model section). We find that Genie generates structures with diverse helices and strands, but FrameDiff- and RFDiffusion-generated structures show greater SSE diversity, particularly for beta sheets.

We also consider tertiary diversity. We first hierarchically cluster generated structures based on pairwise TM scores using single linkage for cluster distances and a TM threshold of 0.6, similar to the approach of RFDiffusion and FrameDiff. Each pairwise TM score is computed using TMAlign \cite{zhang2004scoring} and normalized by the length of the larger protein. We then define tertiary diversity as the number of clusters divided by the number of generated structures. We find that all models are capable of generating diverse structures (Table \ref{table_results} (``Diversity")), with RFDiffusion achieving best performance, followed by Genie (both Genie-SCOPe and Genie-SwissProt) and FrameDiff. When trained with a larger dataset, Genie achieves higher designablity at the cost of reduced diversity, likely due to small model size.

\begin{figure*}[h!]
\begin{center}
\centerline{\includegraphics[width=\textwidth]{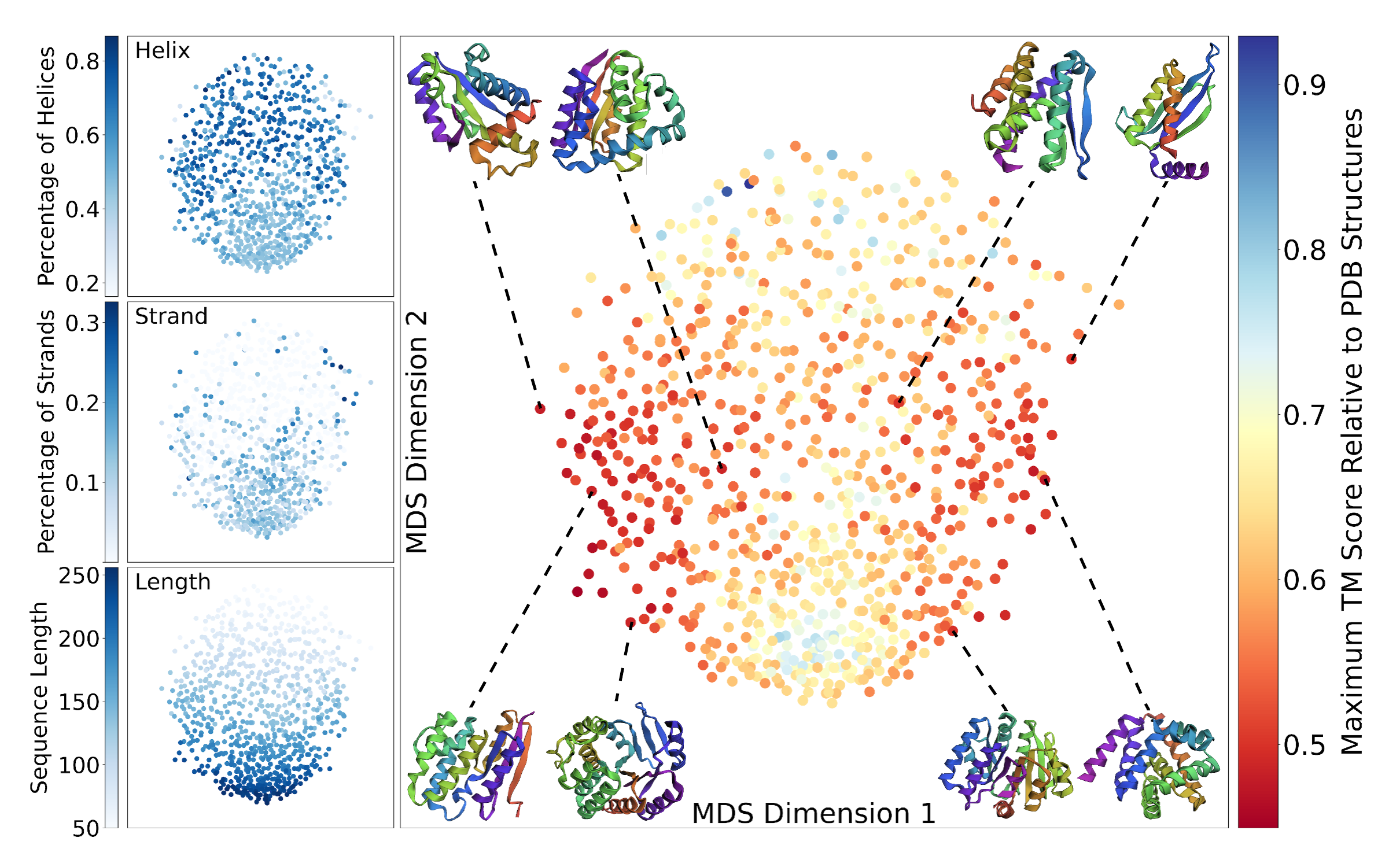}}
\vspace{-0.5cm}
\caption{Design space of Genie-SwissProt. 818 confidently designable structures were embedded in 2D space using multidimensional scaling (MDS) with pairwise TM scores as the distance metric. Domains are colored by their maximum TM score to PDB structures (central panel), fraction of helical residues (top left panel), fraction of beta strand residues (middle left panel), and sequence length (bottom left panel). Eight novel designed domains are shown as representatives. }
\label{novelty_swissprot}
\end{center}
\vskip -0.2in
\end{figure*}

To better assess this tradeoff between designability and diversity, we compute the $F_1$ score as the harmonic mean of designability ($p_{\text{structures}}$) and diversity ($p_{\text{clusters}}$):
\[F_{\beta} = (1 + \beta^2) \cdot \frac{p_{\text{structures}} \cdot p_{\text{clusters}}}{(\beta^2 \cdot p_{\text{structures}}) + p_{\text{clusters}}}\]
where $\beta \in R^+$ tunes the relative importance of designability vs. diversity. We set $\beta=1$ to weigh both equally.

We report $F_1$ scores in Table \ref{table_results} (``$F_1$"). Genie outperforms FrameDiff on both metrics even when trained on the smaller SCOPe dataset while RFDiffusion performs best. However, RFDiffusion is pretrained on predicting protein structures while Genie is trained from scratch. RFDiffusion also contains around 14 times more parameters than Genie (59.8M versus 4.1M). In Appendix \ref{appendix_d3}, we further compute the maximum TM score to other generated structures for every generated structure and provide a visualization of the distribution of these values versus scRMSD.

\subsubsection{Novelty} \label{sec3.2.3}

To quantify the novelty of generated structures, we compute their maximum TM scores with respect to all monomers in the Protein Data Bank (PDB) \cite{berman2000protein}. Following \citeauthor{yim2023se} \yrcite{yim2023se}, we filter the PDB by restricting protein lengths to lie between 60 and 512 residues and resolution to be 5{\AA} or better, which yields 26,468 monomers. To classify a structure as novel, we require that it is confidently designable ($\text{scRMSD} < 2$; $\text{pLDDT} > 70$) and that its maximum TM score to the PDB is less than 0.5. Overall novelty is reported as the percentage of novel structures in Table \ref{table_results} (``Novelty"). Genie generates more novel structures than FrameDiff but RFDiffusion performs best. Appendix \ref{appendix_d4} further visualizes the distribution of maximum TM score to PDB structures versus scRMSD. We note that the novelty results reported here might overestimate Genie-SwissProt's performance due to the larger size of the AlphaFold-SwissProt database relative to the PDB. Nonetheless, the performance of Genie-SCOPe, which does not have this caveat, exhibits the same trend.

\subsubsection{Speed} \label{sec3.2.4}

Because the long models we assess can be computationally expensive, generation speed becomes an important factor. We profile models using one Nvidia A6000 GPU on sequences of length 200 (Table \ref{table_results}). We find FrameDiff to be the fastest, followed by RFDiffusion and Genie. While Genie is the smallest method of the set, it uses more time steps to generate samples and employs triangular multiplicative update layers which have $O(N^3)$ time complexity.

\subsection{Visualization of Design Space}
\label{sec3.3}

To visualize Genie's design space, we apply multidimensional scaling (MDS) to the pairwise TM scores of all 818 confidently designable structures generated by Genie-SwissProt and show the resulting 2D space in Figure \ref{novelty_swissprot}. By overlaying maximum TM scores (relative to PDB structures) and SSE content on the MDS embedding, we confirm that Genie-SwissProt generates diverse structures. We illustrate the quality and diversity of these structures by showing eight novel designs chosen from diverse embedding locations. Appendix \ref{appendix_d5} provides more visualizations of novel structures. We also perform the same analysis for the short version of Genie in Appendix \ref{appendix_c2} and observe similar trends. Interestingly, the short version of Genie generates more diverse structures than Genie-SwissProt.

\section{Conclusion}
\label{conclusion}

In this work we present Genie, a DDPM for \textit{de novo} protein design that substantially outperforms short structure-based models and achieves competitive performance relative to much larger long models. One important contributing factor to Genie's success is the use of dual representations for protein residues. By representing a protein as a sequence of $C_\alpha$ coordinates in Cartesian space instead of FS frames, we can perform diffusion by injecting isotropic Gaussian noise into $C_\alpha$ coordinates, bypassing the need to noise rotation matrices, a more delicate task. On the other hand, during noise prediction, proteins are represented as sequences of FS frames, allowing Genie to reason about inter-residue orientations and achieve high structural quality. Thus Genie simultaneously achieves simplicity of design and geometric expressiveness. Noise prediction is accomplished by combining IPA with backbone updates, which provide a powerful way to reason spatially about protein structure, maintaining equivariance to both translations and rotations while being sensitive to reflections.

Future directions center around three areas. First is scaling up model size, as current versions appear to be capacity-limited. Second is concurrently generating sequence along with structure. Third is conditionally generating structures based on geometric or functional criteria. Such conditional generation has been achieved by other methods, for example in producing structures that contain functional sites \cite{wang2022scaffolding}. The use of pretrained classifiers (\textit{e.g.}, function classifiers) to guide DDPM generation towards novel proteins with desired properties is particularly promising, including in drug discovery. Recent works \cite{Ingraham2022.12.01.518682,Watson2022.12.09.519842,hie2022high} have shown promise in this direction and we hope that the innovations introduced by Genie will further drive progress.

\bibliography{refs}
\bibliographystyle{icml2023}

\newpage
\onecolumn
\appendix

\section{Genie Architectural Details}
\label{appendix_a}

\subsection{Illustration of Protein Backbone Diffusion}
\label{appendix_a0}

\begin{figure*}[h!]
\begin{center}
\centerline{\includegraphics[width=0.9
\textwidth]{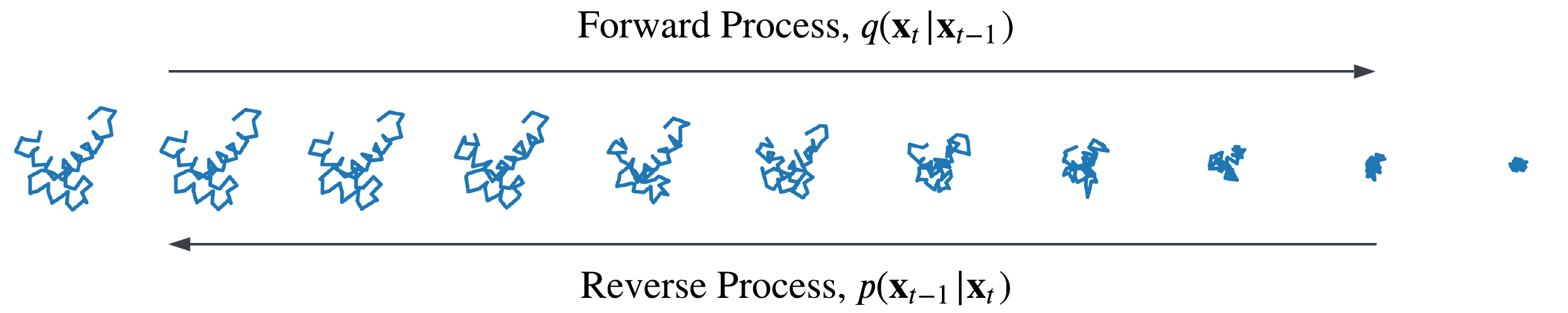}}
\caption{Diffusion of protein backbone in Cartesian space.}
\label{appendix_diffusion}
\end{center}
\vskip -0.2in
\end{figure*}

Figure \ref{appendix_diffusion} visualizes the diffusion process of a protein backbone in Cartesian space. The forward process iteratively adds isotropic Gaussian noise to $C_\alpha$ coordinates, while the reverse process iteratively denoises noisy coordinates through an SE(3)-equivariant model.

\subsection{Encodings}
\label{appendix_a1}

\textbf{Sinusoidal encoding of diffusion step} \quad Let $T$ denote the total number of diffusion steps and $D$ denote the encoding dimension. We define the sinusoidal encoding of diffusion step $t$ as
\[\phi(t) = \begin{bmatrix}
  f(t, 1) \\
  f(t, 2) \\
  \vdots \\
  f(t, D) \\
\end{bmatrix}\]
where
\[ f(t, d) = \begin{cases}
    \sin{\left( t \cdot \pi / T^{\frac{2 \cdot d}{D}} \right)}, &\text{if } d \;\mathrm{mod}\; 2 = 0 \\
    \cos{\left(t \cdot \pi / T^{\frac{2 \cdot (d - 1)}{D}} \right)}, &\text{otherwise}
\end{cases} \]

\textbf{Sinusoidal encoding of residue index} \quad Let $N$ denote the maximum sequence length. We define the sinusoidal encoding of residue index $n$ as
\[\phi(n) = \begin{bmatrix}
  f(n, 1) \\
  f(n, 2) \\
  \vdots \\
  f(n, D) \\
\end{bmatrix}\]
where
\[ f(n, d) = \begin{cases}
    \sin{\left( n \cdot \pi / N^{\frac{2 \cdot d}{D}} \right)}, &\text{if } d \;\mathrm{mod}\; 2 = 0 \\
    \cos{\left(n \cdot \pi / N^{\frac{2 \cdot (d - 1)}{D}} \right)}, &\text{otherwise}
\end{cases} \]

\textbf{Relative positional encoding} \quad We use a linear function to encode the $k$-clipped relative position between residue pairs. Given residue $i$ and residue $j$, we first compute the clipped distance between residues in the amino acid sequence using
\[ d(i, j) = \min(\max(i - j, -k), k) \]
We then compute a one-hot encoding of this clipped distance and use a linear function to project it to the same dimensionality as the pair representation. Residue pairs more than $k$ residues apart are all assigned the same learned encoding.

\subsection{SE(3)-Invariant Encoder}
\label{appendix_a2}

\begin{figure}[h!]
\begin{center}
\centerline{\includegraphics[width=0.9\textwidth]{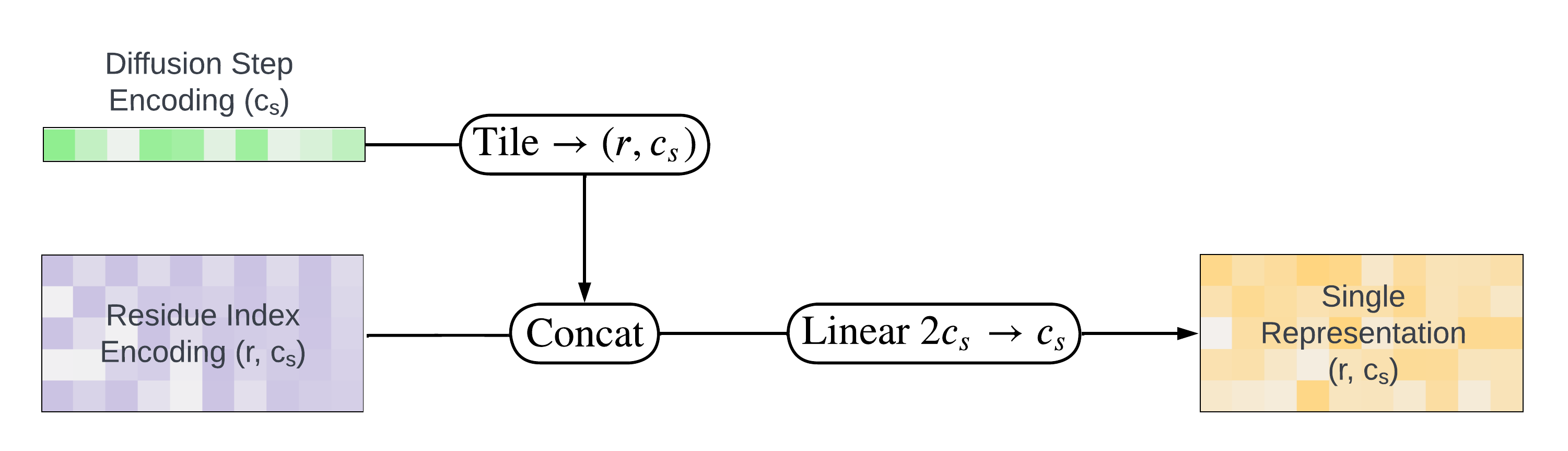}}
\vspace{-0.5cm}
\caption{Architecture of the Single Feature Network, which generates single representations from sinusoidal encodings of the diffusion step and residue index. Notation: $r$: number of residues, $c_s$: dimension of single representation.}
\label{appendix_single}
\end{center}
\vskip -0.2in
\end{figure}

\begin{figure}[ht]
\begin{center}
\centerline{\includegraphics[width=0.9\textwidth]{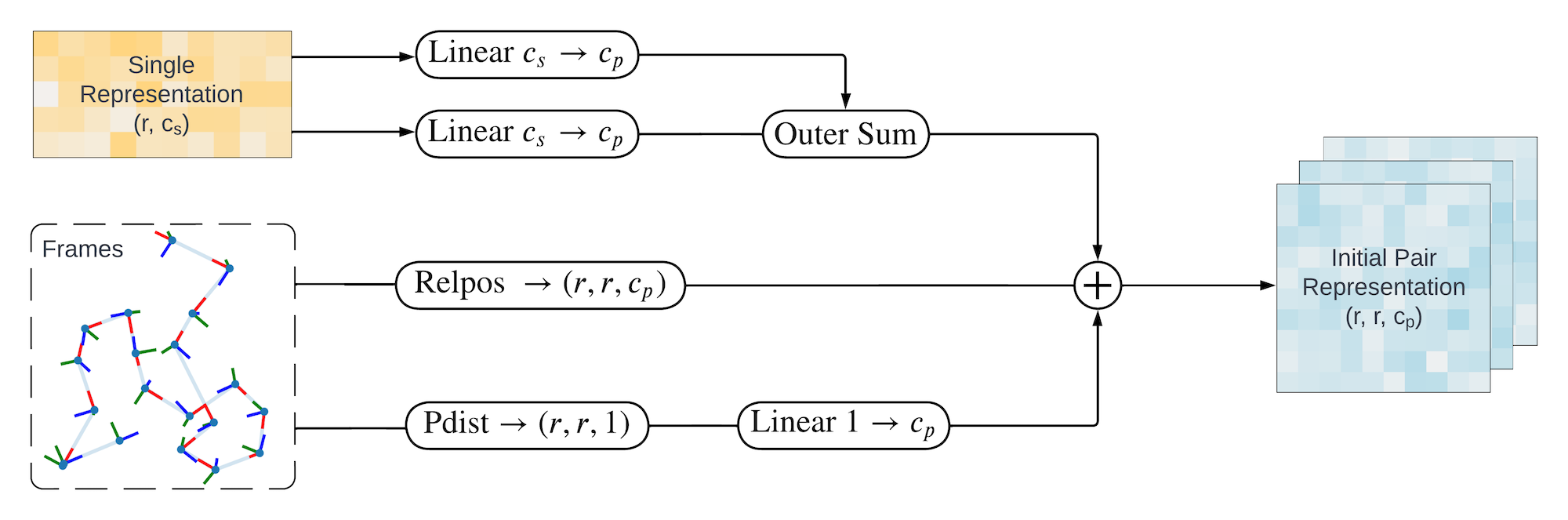}}
\vspace{-0.5cm}
\caption{Architecture of the Pair Feature Network, which generates pair representations from the single representation and Frenet-Serret frames. Notation: $r$: number of residues, $c_s$: dimensionality of single representation, $c_p$: dimensionality of pair representation, $relpos$: relative positional encoding function, $pdist$: pairwise distance function.}
\label{appendix_pair}
\end{center}
\vskip -0.2in
\end{figure}

The SE(3)-invariant encoder constructs and refines single and pair representations based on input FS frames and a sinusoidal encoding of the diffusion step.

\textbf{Single Feature Network} \quad The Single Feature Network constructs the (single) residue representation $\mathbf{s}_i$ by concatenating the sinusoidal encoding of diffusion step $t$ and the sinusoidal encoding of residue index $i$, followed by a linear projection to $\mathbf{s}_i$ (Figure \ref{appendix_single}).

\textbf{Pair Feature Network} \quad The Pair Feature Network computes the pair representation $\mathbf{p}_{ij}$ by summing three latent pair representations -- the relative positional encoding between residue $i$ and $j$ (Appendix \ref{appendix_a1}), the outer summation of the projections of single representations $s_i$ and $s_j$, and the projection of the pairwise distance between residues $i$ and $j$ (Figure \ref{appendix_pair}).

\textbf{Pair Transform Network} \quad The Pair Transform Network uses 5 layers of triangular multiplicative updates (introduced in the evoformer module of AlphaFold2) with pair transitions to refine the pair representations, which are later used by the SE(3)-equivariant decoder to update the single representations and FS frames. We consider a graph over amino acid residues. For each triplet of nodes $ijk$ in the graph, where $i$, $j$ and $k$ are residue indices, edge $ij$ is updated by integrating information from its two adjacent edges $ik$ and $jk$. Since each edge is directed, there are two symmetric updates: the "incoming" edge version which updates edge $ij$ based on the representations of incoming edges $ki$ and $kj$, and the "outgoing" edge version which updates edge $ij$ based on the representations of outgoing edges $ik$ and $jk$. We perform triangular multiplicative updates using both versions.

\subsection{SE(3)-Equivariant Decoder}
\label{appendix_a3}

SE(3)-equivariant decoder utilizes single and pair representations from the previous encoder and refines input frames in a translationally- and rotationally-equivariant manner. Each decoding layer uses Invariant Point Attention (IPA) module to refine single representations and backbone update module to compute and apply frame updates based on updated single representations. In this section, we provide further details on how input frames are refined.

\textbf{Invariant Point Attention} \quad IPA modules computes attention weights by combining standard attention on single representations $s_t$, linear projection of pair representations $p_t$ and squared distance affinities of coordinates in global backbone frames $F_t$. These attention weights are then applied to values computed from single, pair and geometric representations respectively, which are summed to generate updated single representations $s'_t$.

\textbf{Backbone Update Network} \quad The Backbone Update Network utilizes a linear layer to project the updated single representation $(s'_t)^i$ into a translation vector and a rotation matrix (represented as a quaternion with the first component set to 1) for each residue $i$, which are then applied to the corresponding input frame $F_t^i$. Since updates are computed from SE(3)-invariant single representation $s'_t$, arbitrary translation and rotations on input frames $F_t$
 will be preserved in the updated frames $F'_t$
. This implies that the backbone update network is SE(3)-equivariant.

\subsection{Model Hyperparameters}

Table \ref{appendix_encoder_hyperparameters} and Table \ref{appendix_decoder_hyperparameters} summarize the set of hyperparameters for SE(3)-invariant encoder and SE(3)-equivariant decoder respectively.

\begin{table}[htb]
\caption{Hyperparameters for SE(3)-invariant encoder}
\label{appendix_encoder_hyperparameters}
\vskip 0.15in
\begin{center}
\begin{small}
\begin{sc}
\begin{tabular}{lccr}
\toprule
Description & Value \\
\midrule
Dimension of single representations & 128 \\
Dimension of pair representations & 128 \\
Dimension of residue index encoding & 128 \\
Dimension of diffusion time step encoding & 128 \\
Relative positional encoding clipping length & 32 \\
Number of triangular multiplicative update layers & 5 \\
Hidden dimension of triangular multiplicative update layers & 128 \\
Multiplicative factor of pair transition layers & 4 \\
\bottomrule
\end{tabular}
\end{sc}
\end{small}
\end{center}
\vskip -0.1in
\end{table}

\begin{table}[htb]
\caption{Hyperparameters for SE(3)-equivariant decoder}
\label{appendix_decoder_hyperparameters}
\vskip 0.15in
\begin{center}
\begin{small}
\begin{sc}
\begin{tabular}{lccr}
\toprule
Description & Value \\
\midrule
Number of decoding layers & 5 \\
Hidden dimension of IPA & 16 \\
Number of IPA heads & 12 \\
Number of IPA query points & 4 \\
Number of IPA key points & 4 \\
Number of IPA value points & 8 \\
IPA dropout rate & 0.1 \\
\bottomrule
\end{tabular}
\end{sc}
\end{small}
\end{center}
\vskip -0.1in
\end{table}

\subsection{Training}
\label{appendix_a5}

Genie is implemented in PyTorch. For AlphaFold2-inspired components, such as triangular multiplicative updates and Invariant Point Attention, we adapted implementations from OpenFold \cite{Ahdritz2022.11.20.517210}. To train Genie, we use an Adam optimizer with a learning rate of $10^{-4}$. For the shorter version of Genie (that is used for comparisons with ProtDiff and FoldingDiff), we train the model using data parallelism on 2 A100 Nvidia GPUs with an effective batch size of 48. We train Genie for 50,000 epochs ($\sim$9 days). For Genie-SCOPe, we train the model using data parallelism on 12 A100 Nvidia GPUs with an effective batch size of 48. We train Genie-SCOPe for 30,000 epochs ($\sim$2 weeks). For Genie-SwissProt, we train the model using data parallelism on 6 A6000 Nvidia GPUs with an effective batch size of 24. We train Genie-SwissProt for 100 epochs ($\sim$8 days).

\subsection{Evaluation}
\label{appendix_a6}

\begin{figure}[h!]
\begin{center}
\centerline{\includegraphics[width=0.9\textwidth]{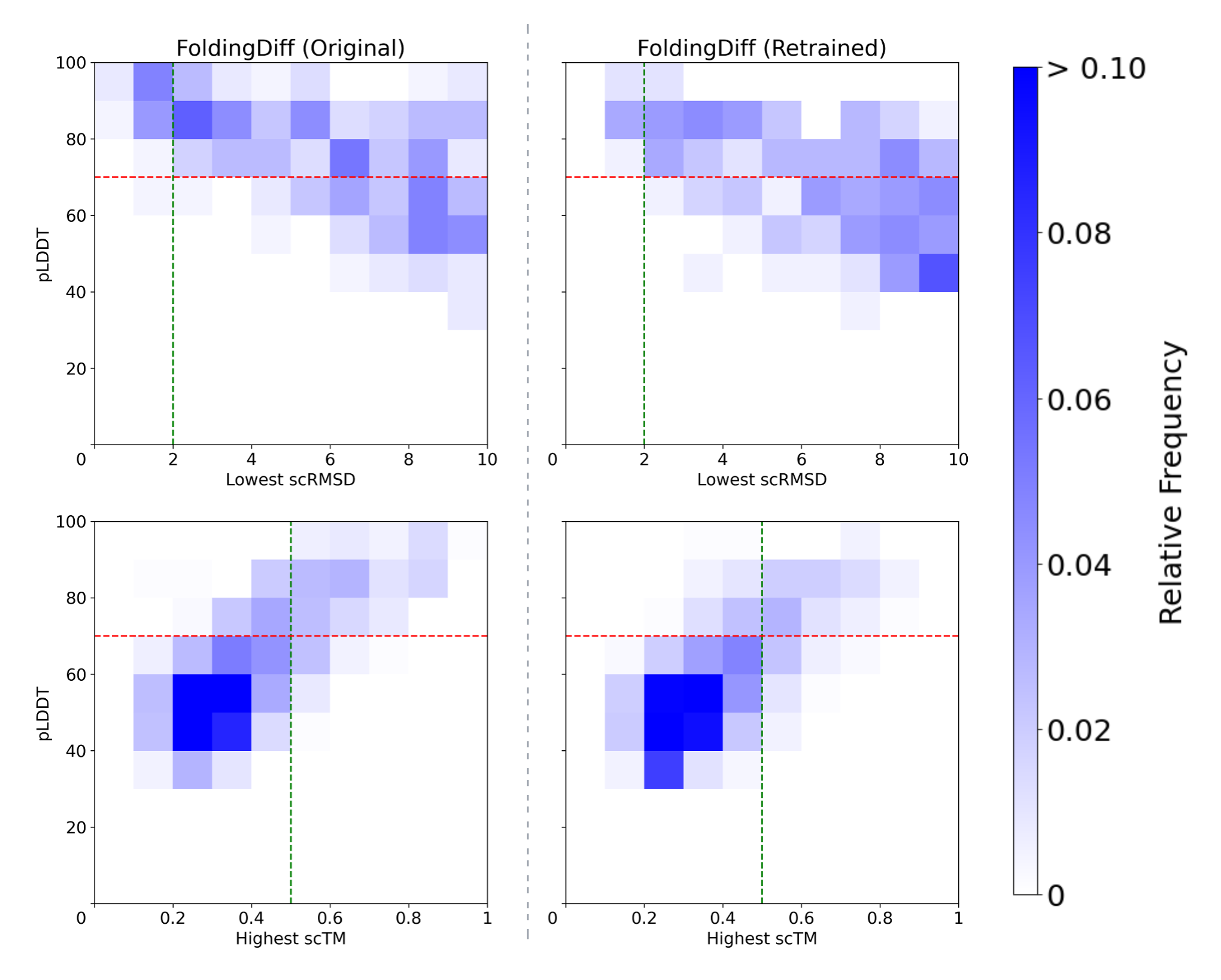}}
\vspace{-0.5cm}
\caption{Comparison on sampling quality for the original and retrained FoldingDiff. The first row compares the distribution of lowest scRMSD versus pLDDT. Higher density in the upper left region indicates higher sampling quality. The second row compares the distribution of highest scTM versus pLDDT. Higher density in the upper right region indicates higher sampling quality. }
\label{appendix_foldingdiff_retrain}
\end{center}
\vskip -0.2in
\end{figure}

For evaluating ProtDiff we retrained the model on our filtered SCOPe dataset (described in \ref{sec2.3}) since the original training dataset and pretrained model weights are not publicly available. The retrained model achieves slightly worse results than the reported model, possibly due to differences in training and dataset size.

For FoldingDiff, the original model is trained using the CATH dataset, filtering at $40\%$ sequence identity and cropping domains longer than 128 residues to 128-residue windows. This leads to 24,316 protein backbones for training. For fair assessment, we retrained the model on our filtered SCOPe dataset and observed degraded performance in the retrained model, mainly due to differences in dataset size. Similar to our results section, for each model, we sample 10 proteins for each sequence length between 50 and 128 residues. Figure \ref{appendix_foldingdiff_retrain} compares the performance of our retrained FoldingDiff and the original FoldingDiff based on the distribution of scRMSD versus pLDDT and the distribution of scTM versus pLDDT. When comparing with Genie in our main results, we continue to use the original version of FoldingDiff (with model weights provided by their paper) for evaluations since it provides better generative performance.

For FrameDiff and RFDiffusion, we use their provided weights for evaluations without any retraining.

\clearpage

\section{Discussion on Self-consistency Pipeline}
\label{appendix_b}

\subsection{Illustration of the Self-consistency Pipeline}
\label{appendix_b1}

\begin{figure}[h!]
\begin{center}
\centerline{\includegraphics[width=0.8\columnwidth]{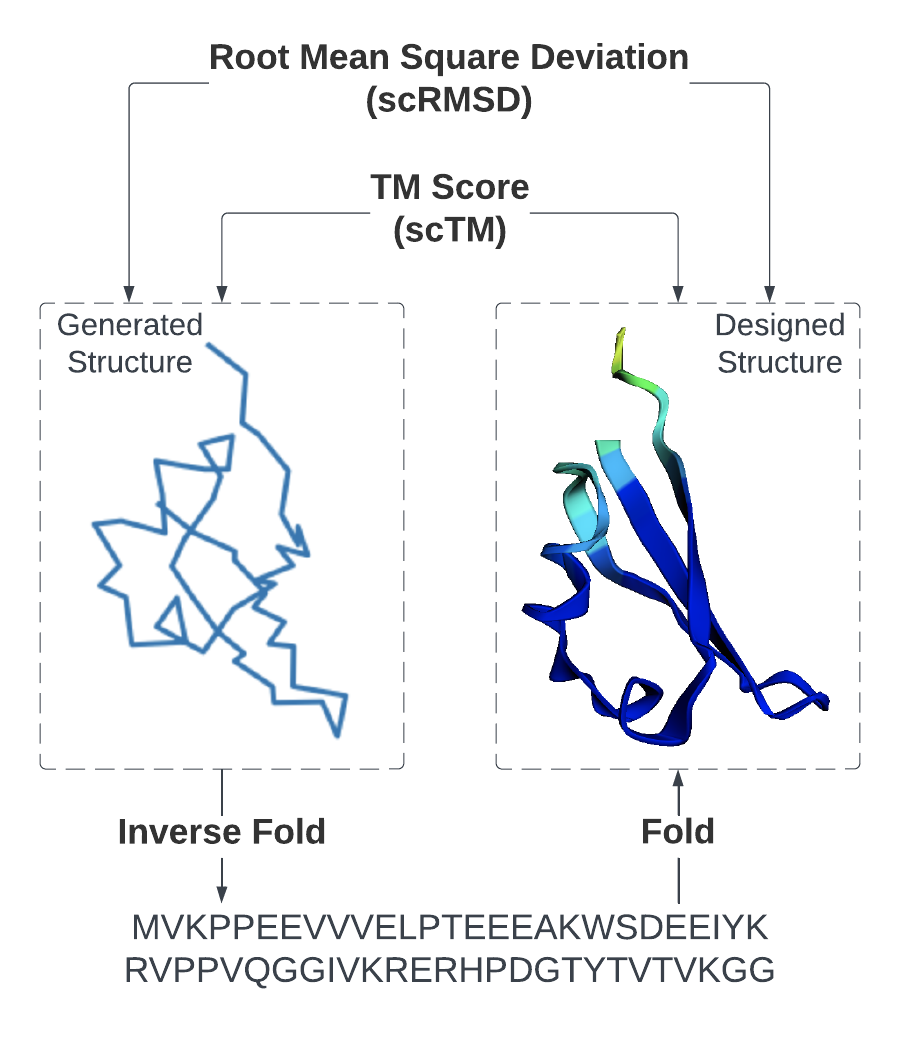}}
\vspace{-0.5cm}
\caption{Self-consistency pipeline. A generated structure is passed through an inverse folding model (\textit{e.g.}, ProteinMPNN) to generate a sequence that is then passed to a protein structure prediction model (\textit{e.g.}, AlphaFold2) to obtain the final designed structure. The self-consistency Template Matching (scTM) score is defined as the TM score between the generated and designed structures, while the self-consistency Root Mean Square Deviation (scRMSD) is defined as the root mean square deviation between the generated and designed structures.}
\label{sctm}
\end{center}
\vskip -0.2in
\end{figure}

Figure \ref{sctm} provides an illustration of the self-consistency pipeline, demonstrating how both scTM and scRMSD are computed given a generated structure.

\newpage

\subsection{Visualization of the Distribution of scTM versus scRMSD}
\label{appendix_b2}

\begin{figure}[h!]
\begin{center}
\centerline{\includegraphics[width=0.9\columnwidth]{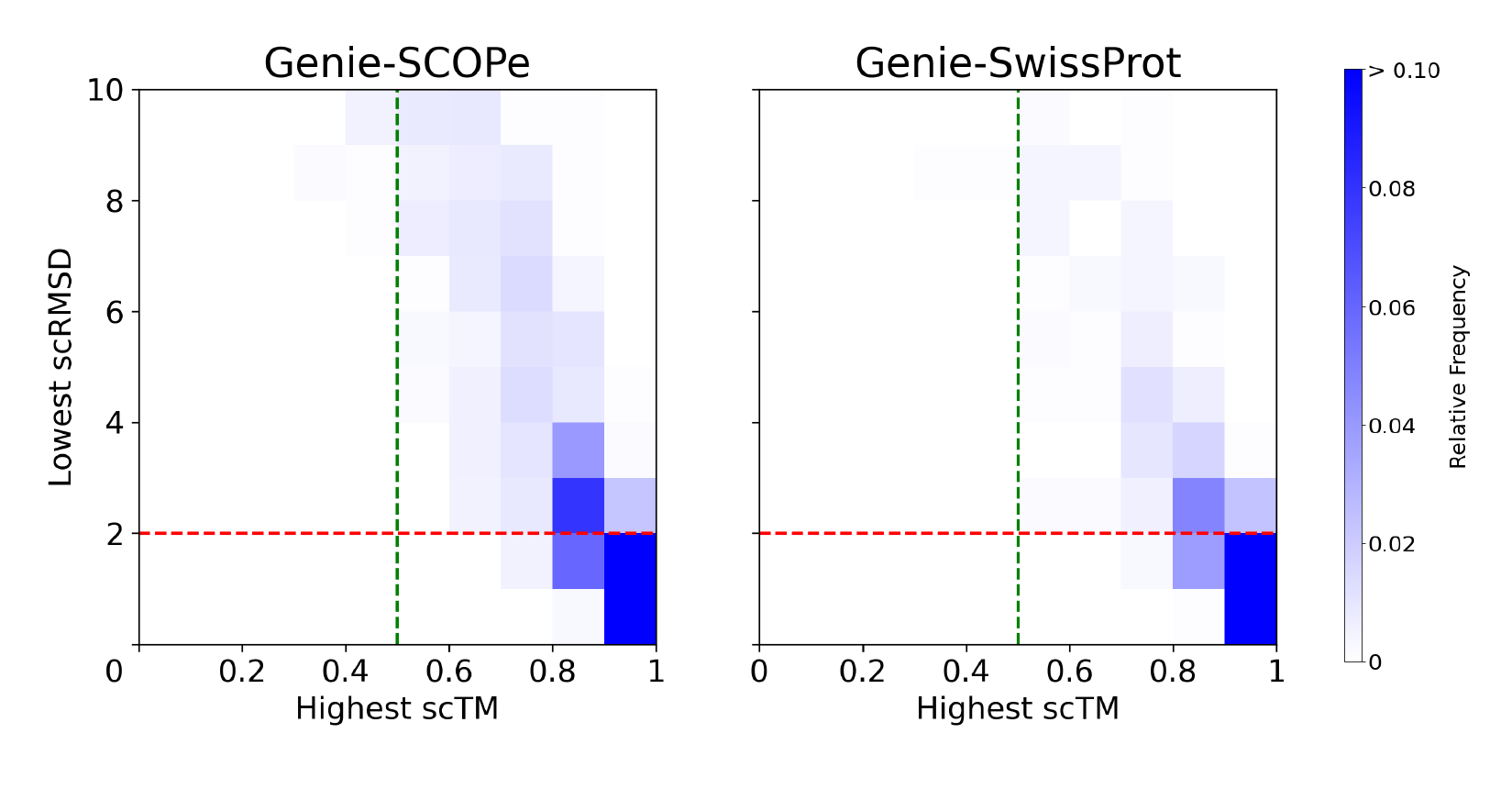}}
\vspace{-0.5cm}
\caption{Distribution of highest scTM versus lowest scRMSD, showing a non-negligible proportion of generated structures with $\text{scTM} > 0.5$ failing to meet the constraint $\text{scRMSD} < 2$. }
\label{appendix_scrmsd_vs_sctm}
\end{center}
\vskip -0.2in
\end{figure}

We visualize the distribution of scTM versus scRMSD for structures generated by Genie-SCOPe and Genie-SwissProt (with maximum sequence length of 256) in Figure \ref{appendix_scrmsd_vs_sctm}. $33.8\%$ Genie-SCOPe-generated structures ($18.3\%$ for Genie-SwissProt) satisfy $\text{scTM} > 0.5$ but have $\text{scRMSD} > 2$, which suggests a mismatch in the generated and designed structures. Conversely, all generated structures with $\text{scRMSD} < 2$ satisfy the constraint $\text{scTM} > 0.5$. This indicates that scRMSD is a more stringent metric for designability compared to scTM.

\newpage

\section{Additional Evaluation Results for Short Models}
\label{appendix_c}

\subsection{Extra Evaluations using ProteinMPNN-ESMFold scTM Pipeline}
\label{appendix_c1}

\begin{figure*}[h!]
\begin{center}
\centerline{\includegraphics[width=\textwidth]{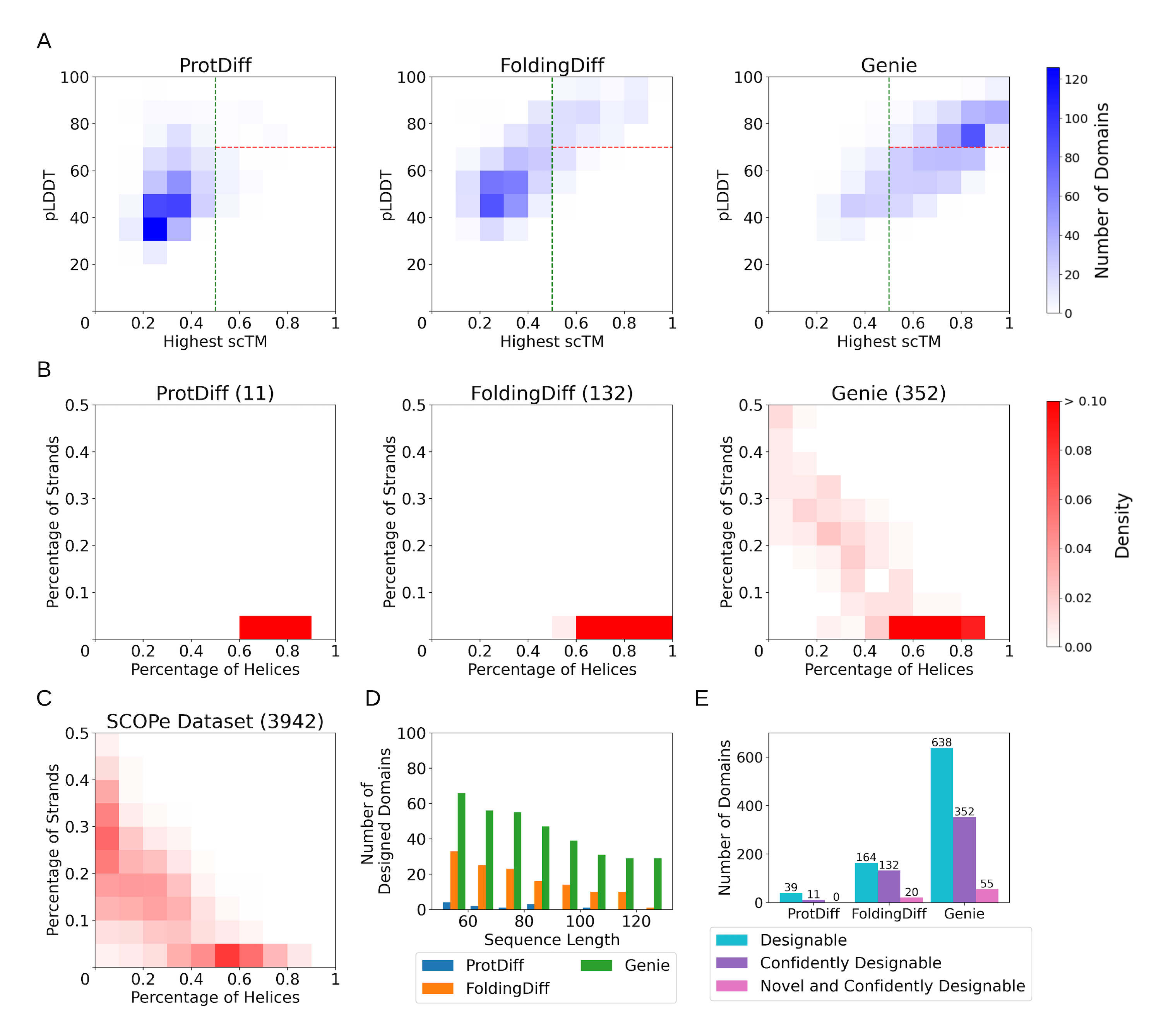}}
\vspace{-0.5cm}
\caption{Evaluation results using ESMFold. (A) Heatmap of the relative frequencies of generated domains with specific combinations of highest scTM and pLDDT values achieved by ProtDiff, FoldingDiff, and Genie. (B) Heatmap of relative frequencies of confidently designable domains with specific combinations of fractional SSE content. The number of designed domains for each model is shown in parentheses. (C) Heatmap of relative frequencies of our SCOPe dataset. This diagram uses the same color scheme as (B) and is provided for reference. (D) Histogram of confidently designable domains as a function of sequence length. (E) Bar chart of number of designable domains generated by different methods out of a fixed budget of 780 attempted designs per method. }
\label{appendix_results}
\end{center}
\vskip -0.2in
\end{figure*}

In Figure \ref{appendix_results} we provide additional evaluations of Genie and other methods by replacing OmegaFold with ESMFold for protein structure prediction in scTM calculations. Overall trends remain unchanged from OmegaFold, except for an overall decrease in the number of confidently designable domains containing $\beta$-strands. 

\newpage

\subsection{Visualization of Genie's Design Space}
\label{appendix_c2}

\begin{figure*}[h!]
\begin{center}
\centerline{\includegraphics[width=\textwidth]{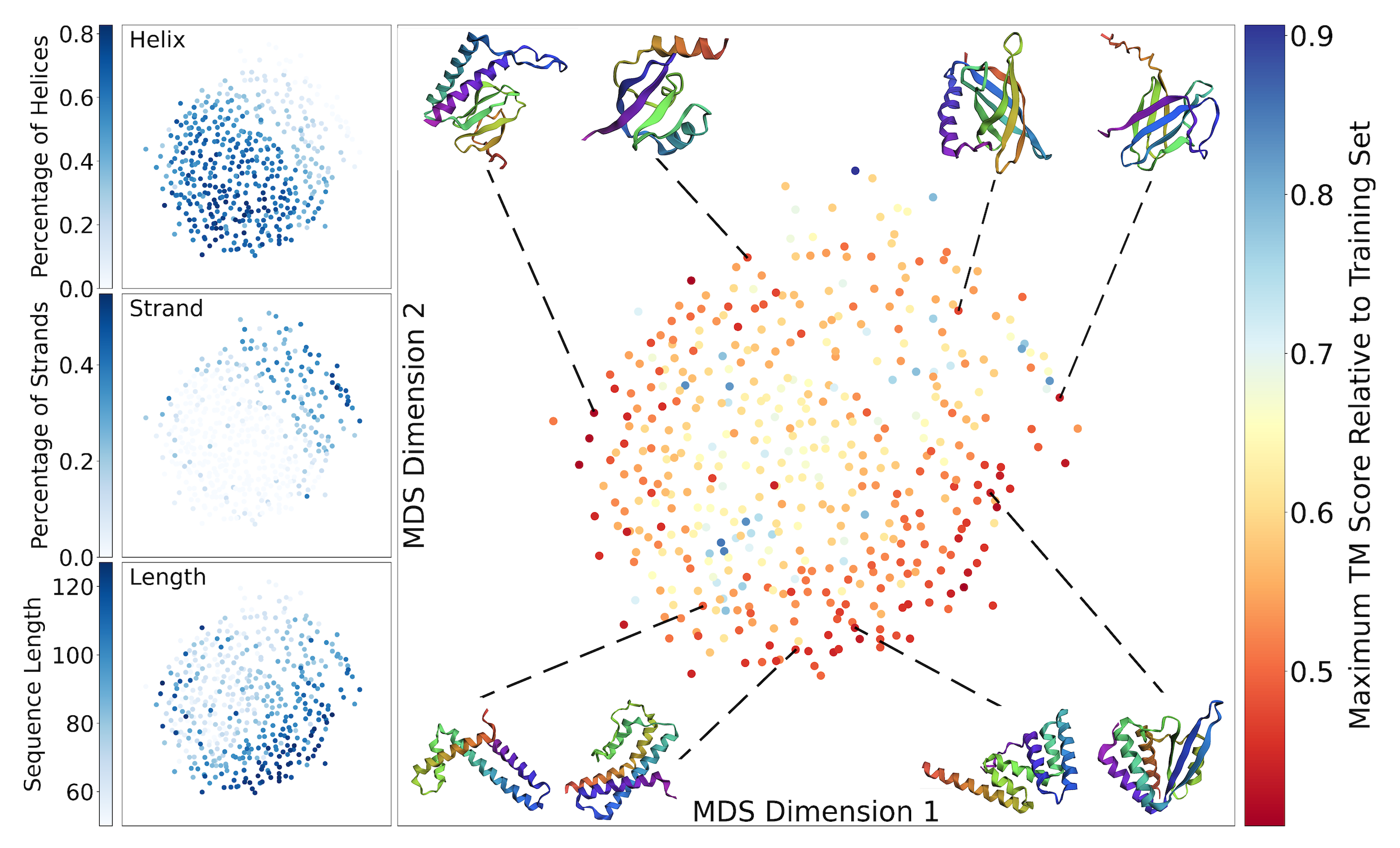}}
\vspace{-0.5cm}
\caption{Design space of Genie. 455 Genie-generated structures that are confidently designable were embedded in 2D space using multidimensional scaling (MDS) with pairwise TM scores as the distance metric. Domains are colored by their maximum TM score to the training set (central panel), fraction of helical residues (top left panel), fraction of beta strand residues (middle left panel), and sequence length (bottom left panel). Eight novel designed domains are shown as representatives. }
\label{appendix_short_novelty}
\end{center}
\vskip -0.2in
\end{figure*}

\newpage

\subsection{Visualizations of Genie-Generated Protein Backbones}
\label{appendix_c3}

\begin{figure}[h!]
\begin{center}
\centerline{\includegraphics[width=\textwidth]{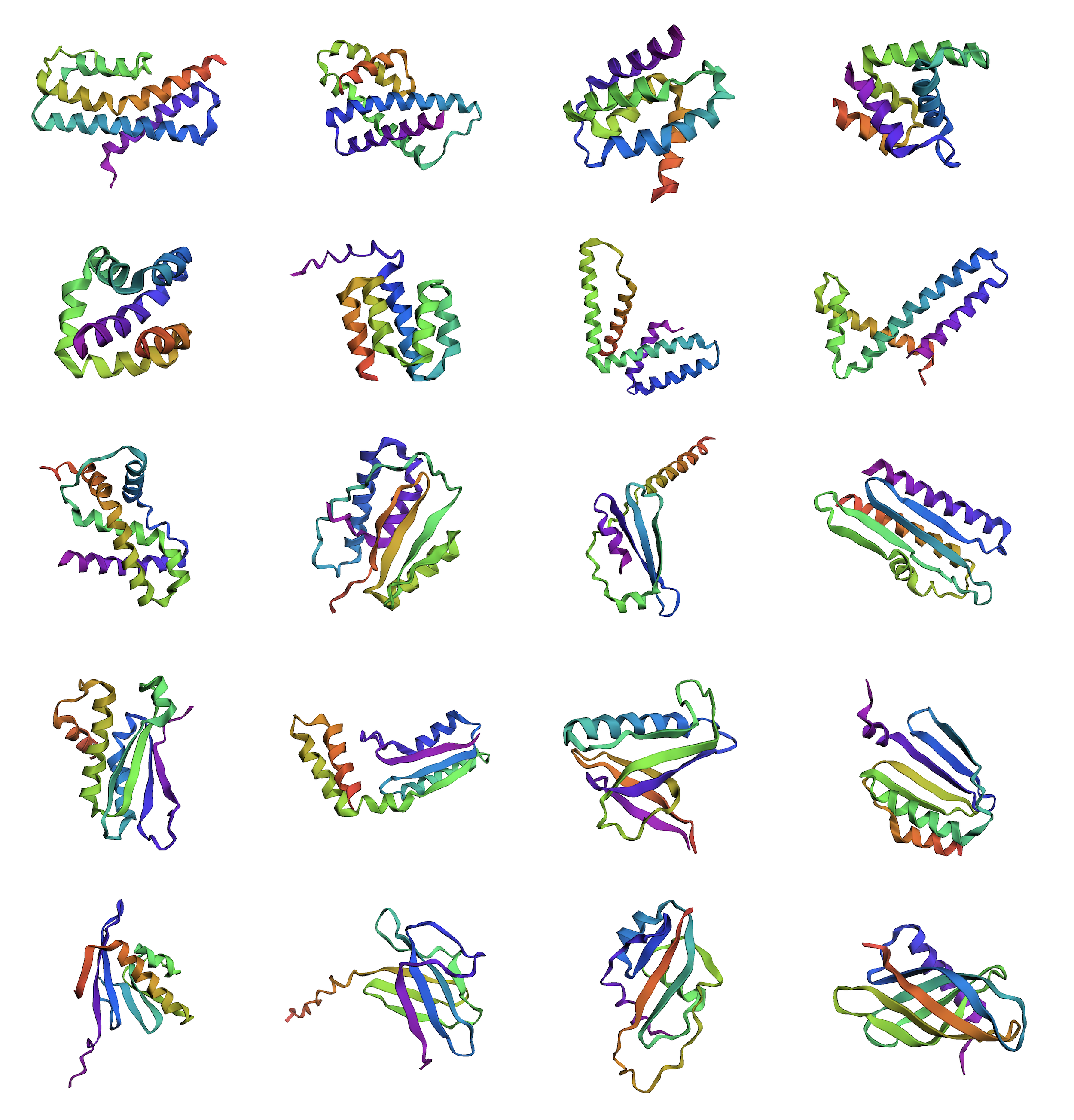}}
\vspace{-0.5cm}
\caption{Additional examples of novel and confidently designable domains generated by Genie.}
\label{appendix_novel}
\end{center}
\vskip -0.2in
\end{figure}

We provide additional visualizations of Genie-generated domains that are novel and confidently designable in Figure \ref{appendix_novel}. 

\clearpage

\section{Additional Evaluation Results for Long Models}
\label{appendix_d}

\subsection{Discussion on the Effect of Sampling Noise Scale}
\label{appendix_d1}

\begin{figure}[h!]
\begin{center}
\centerline{\includegraphics[width=\textwidth]{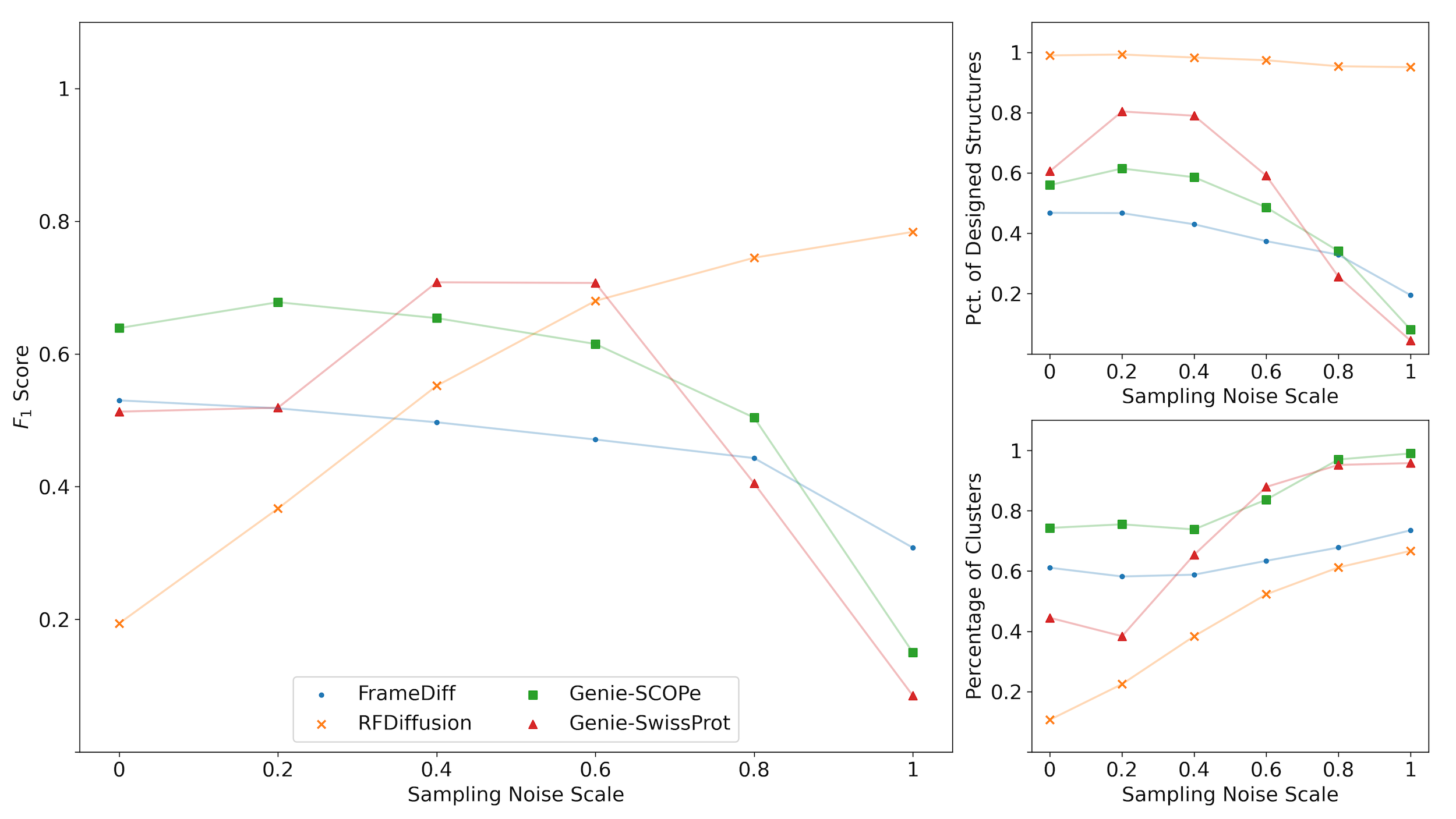}}
\vspace{-0.5cm}
\caption{Additional examples of novel and confidently designable domains generated by Genie.}
\label{appendix_noise_scale}
\end{center}
\vskip -0.2in
\end{figure}

Figure \ref{appendix_noise_scale} shows the effect of sampling noise scale on designability, diversity, and $F_1$ score for RFDiffusion, FrameDiff, and Genie. Generally, lower sampling noise leads to higher quality structures at a cost of lower diversity. With a sampling noise scale of 0.4, Genie achieves an optimal balance between sample quality and sample diversity.

\newpage

\subsection{Visualizations of the Distribution of scRMSD/scTM versus pLDDT for Long Models}
\label{appendix_d2}

\begin{figure}[h!]
\begin{center}
\centerline{\includegraphics[width=\textwidth]{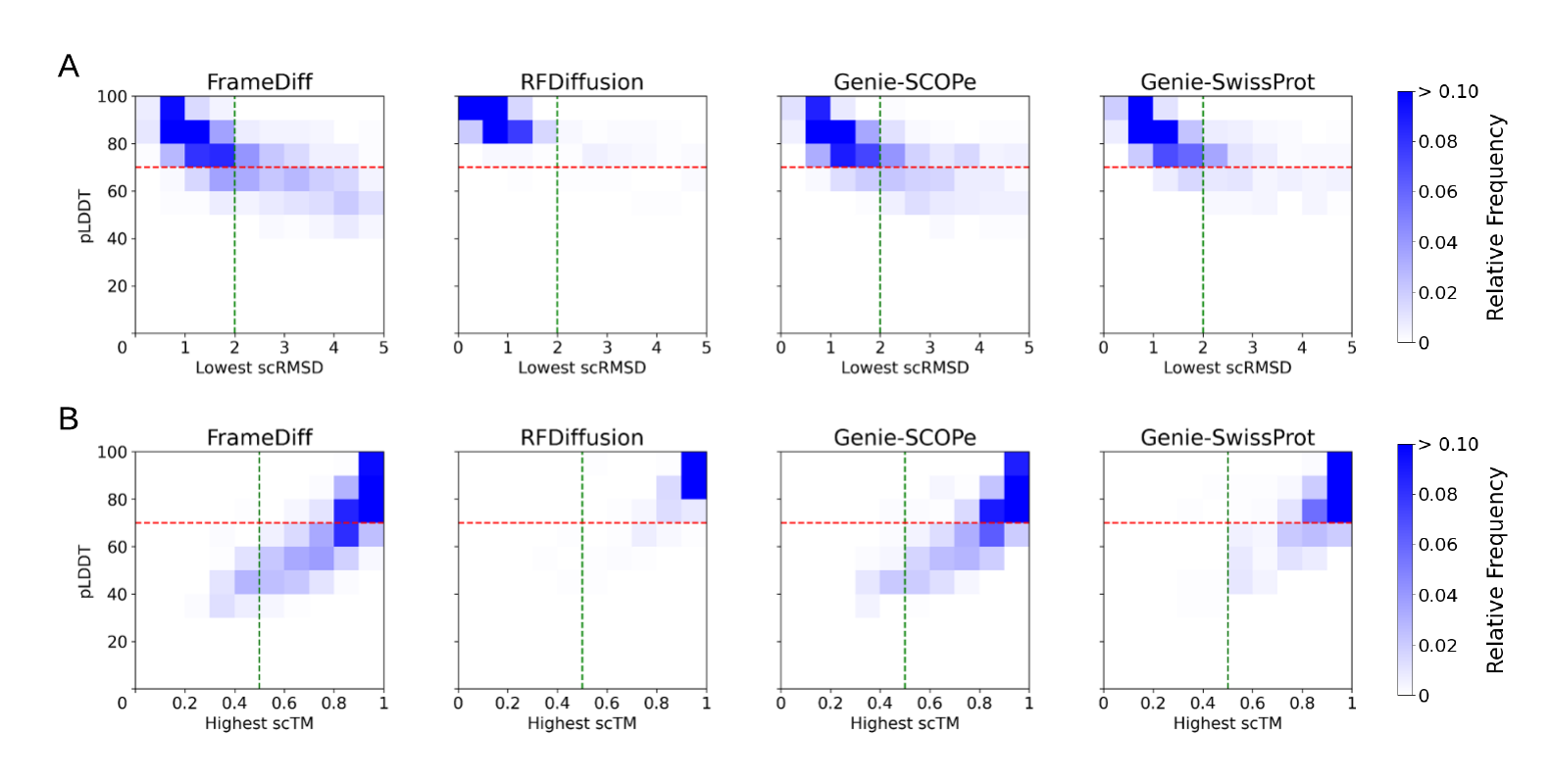}}
\vspace{-0.5cm}
\caption{Visualizations of the distribution of scRMSDs/scTMs versus pLDDTs for long models. (A) Heatmap of relative frequencies of generated structures with specific combinations of lowest scRMSD and pLDDT. Higher density in the top left region indicates higher sampling quality. (B) Heatmap of relative frequencies of generated structures with specific combinations of highest scTM and pLDDT. Higher density in the top right region indicates higher sampling quality. }
\label{appendix_long_sctms}
\end{center}
\vskip -0.2in
\end{figure}

\newpage

\subsection{Visualization of Tertiary Diversity versus Designability}
\label{appendix_d3}

\begin{figure}[h!]
\begin{center}
\centerline{\includegraphics[width=0.9\linewidth]{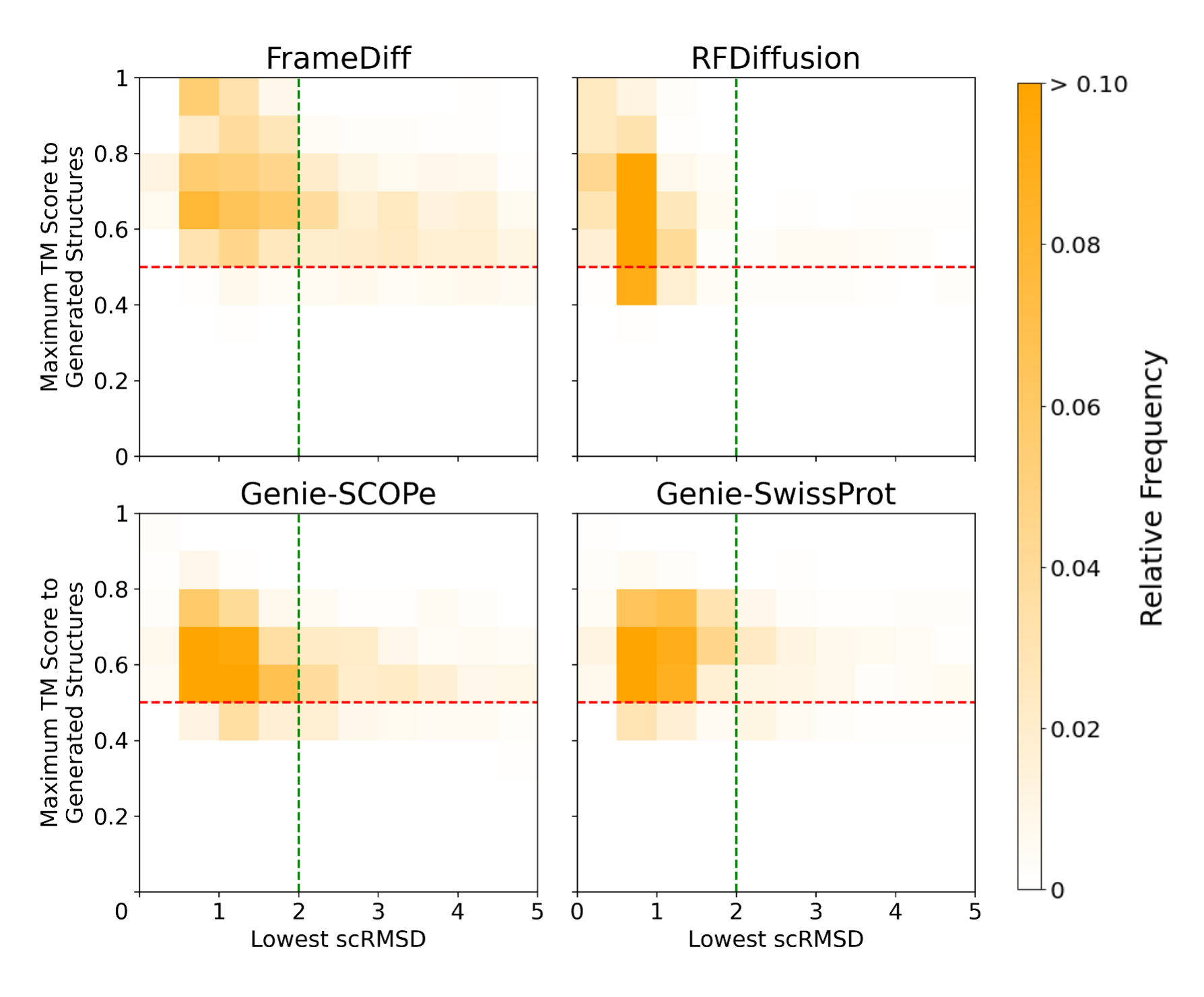}}
\vspace{-0.3cm}
\caption{Heatmap of relative frequencies of generated structures with specific combinations of lowest scRMSD and maximum TM scores to generated structures for Genie and long models. This equivalently shows the density of generated structures on designability versus tertiary diversity. Higher density in the lower left region indicates higher designability and diversity.}
\label{appendix_diversity_vs_quality}
\end{center}
\vskip -0.2in
\end{figure}

\newpage

\subsection{Visualization of Novelty versus Designability}
\label{appendix_d4}

\begin{figure}[h!]
\begin{center}
\centerline{\includegraphics[width=0.9\linewidth]{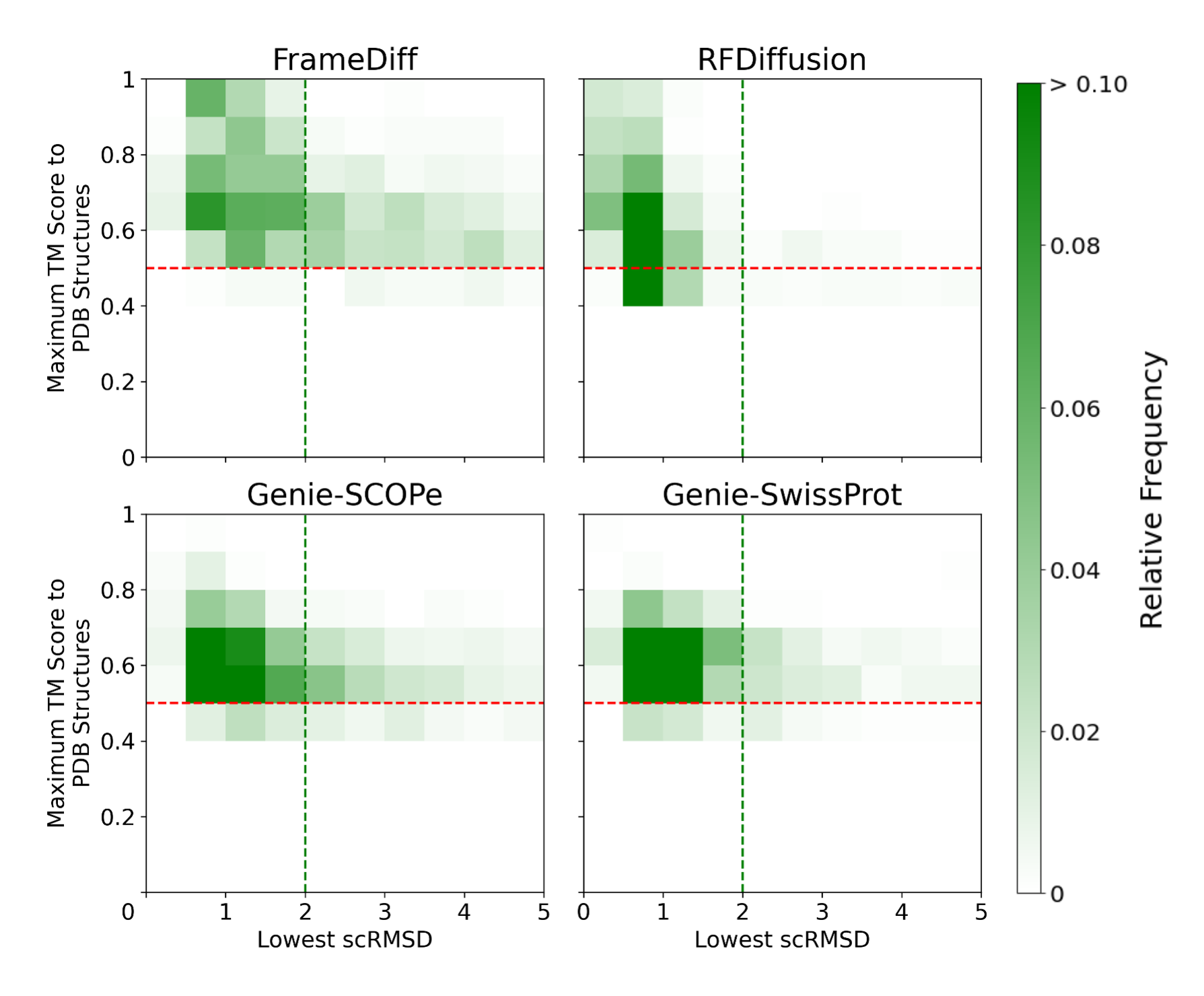}}
\vspace{-0.3cm}
\caption{Heatmap of relative frequencies of generated structures with specific combinations of lowest scRMSD and maximum TM scores to PDB structures for Genie and long models. This equivalently shows the density of generated structures on designability versus novelty. Higher density in the lower left region indicates higher designability and novelty.}
\label{appendix_novelty_vs_quality}
\end{center}
\vskip -0.2in
\end{figure}

\newpage

\subsection{Visualizations of Protein Backbones Generated by Genie-SwissProt}
\label{appendix_d5}

\begin{figure}[h!]
\begin{center}
\centerline{\includegraphics[width=\linewidth]{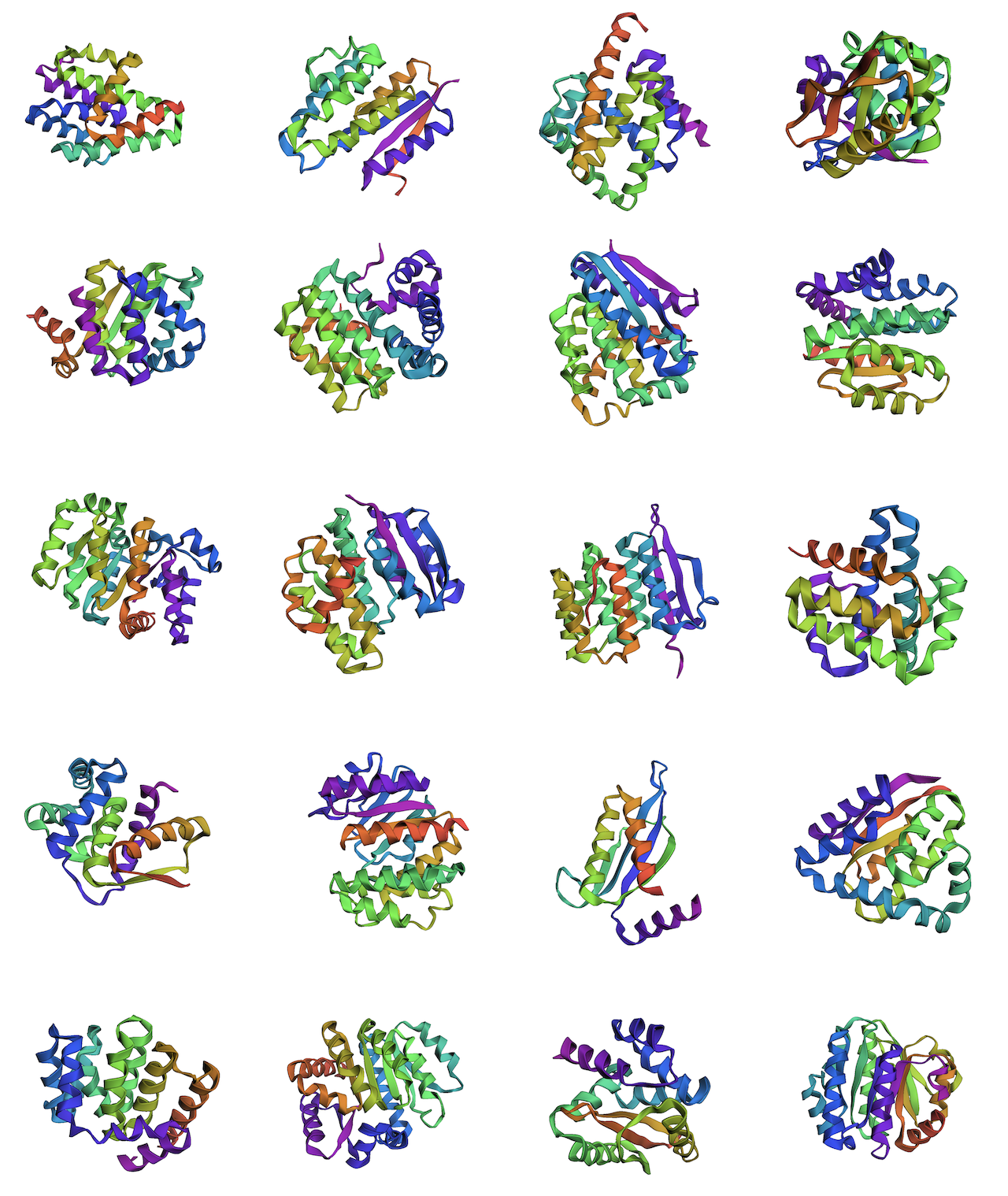}}
\vspace{-0.3cm}
\caption{Additional examples of novel and confidently designable structures generated by Genie-SwissProt.}
\label{appendix_samples_genie_swissprot}
\end{center}
\vskip -0.2in
\end{figure}

\end{document}